\documentclass[conference,compsoc]{IEEEtran}

\usepackage{amsmath}
\usepackage{tikz}
\usepackage{cite}
\usepackage{xurl}
\usepackage{amssymb}
\usepackage[hidelinks]{hyperref}
\usepackage{glossaries-extra}

\usepackage[normalem]{ulem}
\usepackage[inline]{enumitem}
\setlist[enumerate,1]{label=(\arabic*)}
\usepackage{xspace}
\usepackage{listings}
\usepackage{float}
\usepackage{booktabs}
\usepackage{subcaption}
\usepackage{multirow}
\usepackage{amsfonts}
\usepackage{soul}
\usepackage{xcolor}
\usepackage[table]{colortbl}
\usepackage{versions}
\usepackage{bbm}
\usepackage{pifont}
\usepackage[dvipsnames]{xcolor}
\usepackage{tcolorbox}
\tcbuselibrary{listings, skins}
\usepackage{pgfplots}

\usepackage{amsthm}
\usetikzlibrary{patterns}
\pgfplotsset{compat=1.18} 
\usepackage{adjustbox}
\usepackage[ruled,vlined,noresetcount]{algorithm2e}
\usepackage{caption}
\usepackage[lambda,advantage,operators,probability]{cryptocode}

\usepackage[normalem]{ulem}

\lstdefinelanguage{Tamarin}{
  keywords={rule, let, in, builtins, equations, restriction, lemma,
            all-traces, exists-trace, Fr, In, Out, theory, begin, end, pk, All, Ex, not, h, K, reuse},
  keywordstyle=\color{blue!70}\bfseries,
  keywords=[2]{
    SecretKey, Key, QESignKey, QEVerKey,
    InitTD, InitialInstall,
    TDIsRunning, OperationInputs, CompleteOperation, ReportReady, SignatureReady,
    InputMessage, RunOperation,
    TDProducesReport, MacCheck, QEQuotes, Verified, VrfAccepts,
    PrvMemory, ExpectedState, VrfState, VrfSendsChallenge,
    TDMeasurement, ExpectedSoftware, runOp,
    State, PendingLookup, Trained, train, datasets, SecretOutput
  },
  keywordstyle=[2]\color{Bittersweet}\bfseries,
  morecomment=[l]{//},
  morecomment=[l]{///},
  morecomment=[s]{/*}{*/},
  columns=fullflexible,
  keepspaces=true,
  sensitive=true
}

\newcounter{tamarinrule}
\newcounter{tamarinlemma}

\lstdefinestyle{tamarinstyle}{
  language=Tamarin,
  basicstyle=\footnotesize\ttfamily,
  frame=single,
  breaklines=true,
  commentstyle=\color{gray},
  escapechar=|,
mathescape=true,
}

\usetikzlibrary{shapes,arrows}
\usetikzlibrary{backgrounds, positioning, fit}
\makeatletter
\tikzset{
    database/.style={
        path picture={
            \draw (0, 1.5*\database@segmentheight) circle [x radius=\database@radius,y radius=\database@aspectratio*\database@radius];
            \draw (-\database@radius, 0.5*\database@segmentheight) arc [start angle=180,end angle=360,x radius=\database@radius, y radius=\database@aspectratio*\database@radius];
            \draw (-\database@radius,-0.5*\database@segmentheight) arc [start angle=180,end angle=360,x radius=\database@radius, y radius=\database@aspectratio*\database@radius];
            \draw (-\database@radius,1.5*\database@segmentheight) -- ++(0,-3*\database@segmentheight) arc [start angle=180,end angle=360,x radius=\database@radius, y radius=\database@aspectratio*\database@radius] -- ++(0,3*\database@segmentheight);
        },
        minimum width=2*\database@radius + \pgflinewidth,
        minimum height=3*\database@segmentheight + 2*\database@aspectratio*\database@radius + \pgflinewidth,
    },
    database segment height/.store in=\database@segmentheight,
    database radius/.store in=\database@radius,
    database aspect ratio/.store in=\database@aspectratio,
    database segment height=0.1cm,
    database radius=0.25cm,
    database aspect ratio=0.35,
}
\makeatother

\usepackage{environ}

\newcounter{definitionctr}

\NewEnviron{mydefinition}[2][]{%
    \refstepcounter{definitionctr}%
    \ifx\\#1\\\else\label{#1}\fi%
    \begin{figure}[h]%
    \noindent\fbox{%
        \parbox{0.95\columnwidth}{%
        \small
            \textbf{\underline{Definition~\thedefinitionctr:}} #2%
            \vspace{2pt}\par\noindent%
            \BODY%
        }%
    }%
    \end{figure}%
}

\newcommand{\specsize}{\scriptsize}
\newcommand{\code}[1]{{\texttt{\detokenize{#1}}}}
\newcommand{\textcode}[1]{{\small\code{#1}}}

\newcommand{\acron}{\textsf{EVOLVE}\xspace}
\newcommand{\longacron}{\textit{{\textbf{E}nabling \textbf{V}erifiable \textbf{O}utput \textbf{L}ineage for \textbf{V}irtualized ML Workload \textbf{E}xecution}}\xspace}

\newcommand{\prv}{{\ensuremath{\sf{Prv}}}\xspace}
\newcommand{\vrf}{{\ensuremath{\sf{Vrf}}}\xspace}

\newcommand{\chal}{{\textit{Chal}}\xspace}

\usepackage{arydshln}
\newcommand{\app}{{\textsf{App}}\xspace}

\newcommand{\exec}{{\textsf{EXEC}}\xspace}
\newcommand{\execsub}[1]{{\textsf{EXEC}}$_{#1}$\xspace}
\newcommand{\adv}{\ensuremath{\sf{\mathcal Adv}}\xspace}

\newcommand{\vruletab}{\vrule width 0.6pt}

\usepackage{graphicx}
\usepackage{dashbox}
\usepackage{cleveref}

\usetikzlibrary{arrows}
\usetikzlibrary{shapes}

\newcommand{\bverb}[1]{\text{\texttt{#1}}}

\definecolor{darkgreen}{HTML}{00b324}

\newcommand{\Baseline}{white}
\newcommand{\Total}{lightgray}
\newcommand{\BaselineG}{gray}
\newcommand{\TotalG}{black}

\usepgfplotslibrary{groupplots}

\newcounter{protocol}
\newenvironment{protocol}[1]
  {\par\addvspace{\topsep}\noindent
   \begin{tabular}{@{}p{\linewidth}@{}}
   \specialrule{1.2pt}{0pt}{5pt}%
   \leavevmode\refstepcounter{protocol}\textbf{Protocol \theprotocol:} #1 \\
   \specialrule{1.2pt}{4pt}{5pt}}
  { \\
   \specialrule{1.2pt}{4pt}{0pt}
   \end{tabular}\par\addvspace{\topsep}}

\newcommand{\protosep}{\vspace{0.5em}\\ \midrule\vspace{-0.5em}}
\newcommand{\protophase}[1]{\par\vspace{3pt}\noindent\textbf{\textit{#1}}\par\vspace{1pt}}
\newcommand{\protoeq}[1]{\par\vspace{1pt}{\centering$#1$\par}\vspace{1pt}}

\newcounter{contract}
\newcounter{specdef}

\newcommand{\contract}[2]{%
  \par\noindent\refstepcounter{contract}\label{#1}%
  \textbf{\underline{Contract~\thecontract:}}~\textbf{#2}:}

\newcommand{\specdef}[2]{%
  \par\noindent\refstepcounter{specdef}\label{#1}%
  \textbf{\underline{Definition~\thespecdef:}}~#2}

\newenvironment{specblock}
  {\par\vspace{3pt}\noindent\hspace{0.8em}$\begin{aligned}[t]}
  {\end{aligned}$\par\vspace{4pt}}

\makeatletter
\newcommand{\linebreakand}{%
  \end{@IEEEauthorhalign}
  \hfill\mbox{}\par
  \mbox{}\hfill\begin{@IEEEauthorhalign}
}
\makeatother
\glsaddkey
    {hyphenated}        
    {\relax}            
    {\glsentryhyph}     
    {\Glsentryhyph}     
    {\glshyph}          
    {\Glshyph}          
    {\GLShyph}          

\DeclareFontFamily{OT1}{mathc}{}
\DeclareFontShape{OT1}{mathc}{m}{it}{<-> mathc10}{}
\DeclareMathAlphabet{\mathabxcal}{OT1}{mathc}{m}{it}

\setabbreviationstyle[acronym]{long-postshort-user}
\glssetcategoryattribute{acronym}{nohyperfirst}{true}

\newacronym{tee}{TEE}{trusted execution environment}
\newacronym{pox}{PoX}{Proof of execution}
\newacronym{rot}{RoT}{root of trust}
\newacronym{ml}{ML}{machine learning}
\newacronym{zkp}{ZKP}{zero-knowledge proof}
\newacronym{tdx}{TDX}{Trust Domain eXtensions}
\newacronym{td}{TD}{trust domain}
\newacronym{cc}{CC}{Confidential Computing}
\newacronym{gpu}{GPU}{graphics processing unit}
\newacronym{cpu}{CPU}{central processing unit}
\newacronym{ram}{RAM}{random access memory}
\newacronym{npu}{NPU}{neural processing unit}
\newacronym{llm}{LLM}{large language model}
\newacronym{sgx}{SGX}{Software Guard eXtensions}
\newacronym{cvm}{CVM}{confidential virtual machine}
\newacronym{os}{OS}{operating system}
\newacronym{vmm}{VMM}{virtual machine manager}
\newacronym{vm}{VM}{virtual machine}
\newacronym{mac}{MAC}{message authenticated code}
\newacronym{qe}{QE}{quoting enclave}
\newacronym{io}{I/O}{input/output}
\newacronym{spdm}{SPDM}{Security Protocol and Data Model}
\newacronym{peft}{PEFT}{Parameter-Efficient Fine-Tuning}
\newacronym{lora}{LoRA}{Low-Rank Adaptation}
\newacronym{qna}{Q\&A}{question-and-answer}
\newacronym{dcap}{DCAP}{Data Center Attestation Primitives}
\newacronym{nras}{NRAS}{Remote Attestation Service}
\newacronym{awq}{AWQ}{Activation-aware Weight Quantization}
\newacronym{dt}{DT}{decision tree}
\newacronym{mlp}{MLP}{multi-layer perceptron}
\newacronym{tpm}{TPM}{trusted platform module}
\newacronym{ir}{IR}{intermediate representation}
\newacronym{sfi}{SFI}{software fault isolation}
\newacronym{ssa}{SSA}{static single assignment}
\newacronym{elf}{ELF}{executable and linkable format}
\newacronym{ffi}{FFI}{foreign function interface}
\newacronym{tcp}{TCP}{transmission control protocol}
\newacronym{dep}{DEP}{data execution prevention}
\newacronym{ai}{AI}{artificial intelligence}
\newacronym{smpc}{SMPC}{secure multi-party computation}
\newacronym{mcu}{MCU}{microcontroller unit}
\newacronym{ltl}{LTL}{linear temporal logic}
\newacronym{cfi}{CFI}{control-flow integrity}
\newacronym{cet}{CET}{control-flow enforcement technology}
\newacronym{rop}{ROP}{return-oriented programming}
\newacronym{ibt}{IBT}{indirect branch tracking}
\newacronym{dop}{DOP}{data-oriented programming}
\newacronym{ddm}{DDM}{direct data manipulation}
\newacronym{vmpl}{VMPL}{virtual machine privilege level}
\newacronym{sevsnp}{SEV-SNP}{Secure Encrypted Virtualization--Secure Nested Paging}
\newacronym{cca}{CCA}{Confidential Compute Architecture}
\newacronym{rmm}{RMM}{realm management monitor}
\newacronym{pks}{PKS}{protection keys for supervisor}
\newacronym{psp}{PSP}{platform security processor}
\newacronym{sdk}{SDK}{software development kit}
\newacronym{loc}{LOC}{lines of code}
\newacronym{bti}{BTI}{Branch Target Identification}
\newacronym{pac}{PAC}{Pointer Authentication Codes}

\glsdisablehyper

\newif\ifpublicversion
\newif\ifanonversion
\publicversiontrue 

\begin{document}

\title{Enabling Verifiable Output Lineage for Virtualized ML Workload Execution}

\ifpublicversion
\author{
\IEEEauthorblockN{Prach Chantasantitam\textsuperscript{*},
Adam Caulfield\textsuperscript{*},
Sujaya Maiyya\textsuperscript{*},
N. Asokan\textsuperscript{*,\dag}}
\IEEEauthorblockA{\textsuperscript{*}University of Waterloo, \textsuperscript{\dag}KTH Royal Institute of Technology}
\IEEEauthorblockA{\texttt{\{pchantas,acaulfield,smaiyya\}@uwaterloo.ca}, \texttt{asokan@acm.org}}
}
\fi
\ifanonversion
\author{\textit{Anonymous Authors}}
\fi

\maketitle

\begin{abstract}
Machine learning (ML) property attestation lets a model provider demonstrate properties of their models and datasets to verifiers, and trusted execution environments (TEEs) have emerged as its most practical mechanism. However, existing TEE-based systems attest only a static snapshot of the TEE's initial configuration, leaving run-time behavior unauditable. Proof of execution (PoX) architectures close this gap by requiring the prover's root of trust to also demonstrate that (1) the attested operation actually executed and (2) the claimed outputs were affected only by that operation. Existing PoX designs, though, target microcontrollers and rely on hardware features unavailable in server-grade TEEs, so they cannot be applied directly to ML workloads.

We present \acron, a software-based PoX framework for server-grade TEEs. \acron mediates an operation's writes and control flow to track whether a requested ML operation executes without violation. It binds the execution status to measurements of the operation's inputs and outputs, enabling verifiers to assess run-time behavior through ML property attestation. We implement \acron on Intel TDX as a Rust module within its virtual firmware, with policy checks built on primitives formally verified in F*. 
Our evaluation shows that \acron is versatile across diverse ML applications on unmodified hardware, provides formal guarantees for its policy checks. \acron incurs per-request run-time overhead on average ranging from 1.07$\times$ (for decision trees) to 2.13$\times$ (for multi-layer perceptrons).
\end{abstract}


\section{Introduction}


\Gls{ml} property attestation~\cite{duddu2024laminator,chantasantitam2026palm,duddu2024attesting} lets a model provider demonstrate properties of their \gls{ml} assets (models, datasets) to a verifier.
While alternatives such as \glspl{zkp} incur substantial overhead~\cite{zkpot1,zkpot2,zkcnn,zklora,zkllm,zkmlaas} (e.g., 518--4200$\times$ the training time for logistic regression~\cite{zkpot1}) and proof-of-learning schemes can be subverted~\cite{jian_liu_break_PoL,papernot_break_PoL}, \gls{tee}-based approaches~\cite{duddu2024laminator,chantasantitam2026palm,atlas} offer strong security guarantees with reasonable performance.

\Gls{tee}-based \gls{ml} property attestation~\cite{duddu2024laminator,chantasantitam2026palm,exclavefl,atlas} targets server-grade machines, building on Intel \gls{sgx}~\cite{sgx}, Intel \gls{tdx}~\cite{tdx}, or AMD \gls{sevsnp}~\cite{amd-sev-snp}. These technologies protect against an untrusted host \gls{os} or \gls{vmm} that retains higher privilege and mediates the \gls{tee}'s access to resources and execution. Their attestation support produces authenticated evidence of the \gls{tee}'s configuration, which a verifier requests from a \gls{tee}-based prover in remote attestation. 
For \gls{ml} property attestation,
code running inside the \gls{tee} hashes an operation's inputs and outputs
and includes those hashes in the attestation evidence,
binding them to the \gls{tee}'s initial configuration~\cite{duddu2024laminator,atlas,exclavefl,chantasantitam2026palm}.

Because this snapshot captures only the \gls{tee}'s initial configuration and auxiliary data, attacks that alter run-time behavior within the \gls{tee} go undetected. Such attacks are far from hypothetical: run-time vulnerabilities are prevalent in widely used \gls{ml} libraries (e.g., PyTorch~\cite{pytorch}, TensorFlow~\cite{tensorflow}, and \texttt{pickle}-based loading~\cite{model_code_backdoors,sleepy_pickle,code_poisoning_attacks}), as recent disclosures show~\cite{cve-2025-32434,cve-2026-24747}. An adversary may corrupt or substitute model weights after measurement~\cite{cve-2025-53630,cve-2025-8747}, or manipulate the execution path so outputs diverge from those produced by the attested model~\cite{cve-2022-45907,cve-2025-32444}. Another example is corruption of \textit{model routing}, which selects among collocated models at run-time; recent work shows that routing decisions can be manipulated without modifying the underlying models or code~\cite{shafran2025rerouting,lin2025life,tang2026route,jiang2026sparse}.
Because attestation evidence in these systems is inherently \emph{static}, it alone cannot reveal run-time deviations occurring within \gls{tee}.

\gls{pox} architectures~\cite{asap,apex,pearts,slapp,flicker} extend remote attestation to run-time properties: a root of trust in the prover must also demonstrate that (1) the attested operation was executed, and (2) reported outputs were influenced only by the operation's most recent execution. \gls{pox} has been explored primarily in embedded systems, relying on lightweight hardware extensions to microcontrollers~\cite{asap,apex} or microcontroller-specific features~\cite{pearts,slapp}.

These designs do not transfer directly to server-grade machines for three reasons. (1) \gls{pox} requires a trusted mechanism that detects whether any code other than the operation itself has written to the operation's output.
Prior approaches implement this with custom hardware or the \gls{tee}'s isolation boundary. In \gls{cvm}-based \glspl{tee}, however, that boundary encloses the entire VM, so using it to separate operations would require one \gls{cvm} per operation (e.g., as in~\cite{exclavefl,attestable_audits}).
(2) the \gls{pox} root of trust must be the only component that can update the \gls{pox} data, which requires hardware support to protect the metadata from the attested software. Server-grade \glspl{tee} lack intra-\gls{tee} access control, and relying on the guest \gls{os} is unsafe because it may itself be compromised by the run-time attacks that \gls{pox} aims to detect.
(3) Recent work provides part of this missing isolation by building compartments \emph{within} \glspl{cvm}, but relies on platform-specific features, such as \glspl{vmpl} on AMD \gls{sevsnp}~\cite{veil,nestedsgx,cabin,sabanic2026wallet,huang2026exia,exclavefl}, \gls{pks} on Intel \gls{tdx}~\cite{erebor}, Arm \gls{cca}'s granule protection tables~\cite{shelter},
or on software outside the \gls{tee}~\cite{tetd,shi2025serverless}. These designs also isolate whole compartments rather than individual operations, and attest only the code loaded into each, which fails to establish that a given operation was executed or that its outputs reflect only its most recent execution (discussed further in Section~\ref{sec:related_work}).

\gls{ml} workloads executing in \glspl{cvm} especially make \gls{pox} challenging due to the use of shared library code, such as training and inference frameworks. Thus, a \gls{pox} mechanism must be able to provide operation-level \gls{pox} while allowing operations to share libraries.

To address these challenges, we introduce \acron, a software framework for \longacron that enables \gls{pox} for \gls{ml} operations on server-grade \glspl{tee}. \acron consists of an instrumentor and a run-time monitor, addressing the above challenges as follows:
\begin{enumerate}[nosep,leftmargin=*]
    \item The instrumentor routes every write and indirect call in the \emph{payload} (i.e., the \gls{ml} application inside the \gls{cvm}, including its library code) through the monitor, which checks whether the caller belongs to the requested operation.
    \item The monitor runs in the \gls{cvm}'s guest virtual firmware and blocks all payload writes to \gls{pox} data, so only the monitor can record whether the requested operation executed without violations. 
    \item \acron relies only on software instrumentation, a virtual firmware, and standard remote attestation, not on platform-specific features.
\end{enumerate}
As a proof of concept, we implement a prototype on Intel \gls{tdx}, with the instrumentor as an LLVM pass and the monitor as a Rust layer in TD-Shim~\cite{TD-Shim-spec} whose policy checks are built on primitives formally verified in F*~\cite{fstar}.
%
%
In summary, we make the following contributions:
\begin{enumerate}[nosep,leftmargin=*]
    \item We identify challenges of \gls{pox} in \glspl{cvm} and propose \acron, the first software-based framework for operation-level \gls{pox} in \glspl{cvm}. \acron mediates all memory writes and indirect calls to provide remote verifiers with evidence of output provenance (Section~\ref{sec:problem});
    \item We design \acron as two modules: 
    \begin{enumerate}[nosep,label=(\roman*)]
    \item an \emph{instrumentation module} that mediates memory writes and indirect calls with dynamically-defined targets in \gls{ml} operations (Section~\ref{sec:design:inst}), and
    \item a \emph{run-time monitor} extension to the \gls{cvm}'s guest virtual firmware that maintains \gls{pox} data and uses formally verified primitives to enforce access-control and prepare \gls{pox} evidence (Section~\ref{sec:design:monitor});
    \end{enumerate}
    \item We implement and evaluate \acron as a Rust+F* extension to TD-Shim virtual firmware~\cite{TD-Shim-spec} on Intel~\gls{tdx}. We show \acron provides \emph{formal guarantees}, is \emph{versatile} across \gls{cvm} architectures, and easily integrates with several model-routing applications. Based on example \gls{dt} and \gls{mlp} classifiers, its average per-request run-time overheads are 1.07$\times$ ($\pm$0.04) for \gls{dt} and 2.13$\times$ ($\pm$0.81) for \gls{mlp} (Section~\ref{sec:evaluation}). 
\end{enumerate}

\section{Background}\label{sec:background}

\textbf{TEEs and Attestation.}
\glspl{tee} provide isolated execution environments that prevent software outside their trusted boundary from directly accessing protected data. The boundary varies by architecture: Intel \gls{sgx}~\cite{sgx} isolates user-space \textit{enclaves}, while Arm TrustZone~\cite{trustzone} isolates system resources, including a trusted \gls{os}, applications, and peripherals. Server-grade \glspl{tee} also support \glspl{cvm}, which extend this protection to entire \glspl{vm}, as in AMD \gls{sevsnp}~\cite{amd-sev-snp}, Intel \gls{tdx}~\cite{tdx}, and Arm Realms~\cite{realms}. Remote attestation lets a prover demonstrate its identity and configuration to a verifier using authenticated evidence, typically a \gls{mac} or digital signature, generated by a hardware root of trust that binds measurements of the \gls{tee}'s initial state to auxiliary data supplied at run-time. The verifier checks this evidence to determine whether the \gls{tee} is in an expected state.

In Intel \gls{tdx}, \glspl{cvm} are called \glspl{td}. When the untrusted \gls{vmm} creates a \gls{td}, the \gls{tdx} module measures its components. The \gls{td} requests a \texttt{TDREPORT} to attest its state. The \texttt{TDREPORT} contains a \gls{mac} that the \gls{cpu} computes over the \gls{td} image, the \gls{tdx} module image, run-time data in the run-time measurement registers (\texttt{RTMR}s), and \texttt{REPORTDATA}. The \gls{qe}, a dedicated Intel \gls{sgx} enclave~\cite{intel_tdx_base_spec}, verifies the \gls{mac} and generates an \gls{sgx}-style \texttt{QUOTE} signed with a provisioning certification key~\cite{intel_pcs}.
A remote verifier checks the \texttt{QUOTE} to confirm that the expected \gls{td} image produced the \texttt{RTMR}s and \texttt{REPORTDATA} within.

\begin{figure}[t]
    \centering
    \resizebox{\columnwidth}{!}{
    \begin{tikzpicture}
        \fill[blue!40!white] (.75,0) rectangle (1,2.5);
        \fill[blue!40!white] (3.75,0) rectangle (4,2.5);

        \node[text width=3cm] at (2.35,2.75) {\vrf};
        \node[text width=3cm] at (5.15,2.75) {\prv};
        
        \draw[->, thick] (1.25, 2) -- node[above] {\small (1) \chal} (3.5,2);

        \node[right,text width=3cm] at (4,2) {\small (2) Execute $MEM$ \& obtain $\{O, EXEC\}$ };
       
        \node[right,text width=3cm] at (4,.75) {\small (3) Authenticated measurement: $H$};

        \draw[->, thick] (3.5, 0.25) -- node[above] {\small (4) $\{O, EXEC, H\}$} (1.25, 0.25);     
        
        \node[left,text width=3cm] at (2.5,0.25) {\small (5) Verify};
    \end{tikzpicture}
    }
    \vspace{-0.5em}
    \caption{Proof of execution (PoX) built atop traditional remote attestation: a prover (\prv) demonstrates integrity of its software and execution properties to a verifier (\vrf).}
    \label{fig:PoX}
\end{figure}

\textbf{Proofs of Execution.} Existing \gls{tee}-based \gls{ml} property attestation frameworks attest at the granularity of the entire \gls{tee} (e.g., a snapshot of a \gls{sgx} enclave~\cite{duddu2024laminator} or \gls{tdx} \gls{cvm}~\cite{chantasantitam2026palm}). In contrast, \gls{pox} architectures~\cite{asap,apex,pearts,slapp,flicker} extend remote attestation to the execution of individual operations, enabling a prover (\prv) to demonstrate to a verifier (\vrf) that:
\begin{enumerate*}[leftmargin=*,nosep=*]
    \item \prv has a specified operation installed,
    \item \prv executed that operation at the time of the request, and
    \item any outputs \prv reports were modified only by the most recent execution of that operation.
\end{enumerate*}
A typical \gls{pox} protocol (Figure~\ref{fig:PoX}, abstracted from prior work~\cite{apex}) proceeds as follows. \vrf requests \gls{pox} for expected application software (\app) by sending a challenge (\chal), typically a cryptographic nonce. A \gls{rot} in \prv then takes an authenticated measurement over the memory region ($MEM$) housing \app, the outputs ($O$) of \app's execution, execution-status metadata (\exec), and \chal, producing an authentication token ($H$), typically a \gls{mac} or digital signature. \prv returns $O$, \exec, and $H$, which \vrf inspects to decide whether the execution and outputs are trustworthy.

To reflect execution integrity, \gls{pox} requires a \gls{rot} that has exclusive access to \exec, can monitor entry/exit of \app, and can monitor access to $O$, the latter two being how it updates \exec. Prior work meets these requirements with trusted hardware, such as dedicated hardware modules~\cite{asap,apex} or \glspl{tee}~\cite{slapp,pearts}; computing $H$ also requires a secret key, implying a hardware-supported \gls{rot}. \gls{pox} has been explored primarily in embedded systems~\cite{asap,apex,slapp,pearts}, with similar notions on high-end systems relying on architecture-specific capabilities (e.g., late launch in AMD64~\cite{amd_vt}) and \glspl{tpm}~\cite{flicker}.

\section{Problem Statement}\label{sec:problem}



\subsection{Challenges}

Realizing \gls{pox} for \gls{ml} operations requires extending operation-level \gls{pox} from embedded systems to \glspl{cvm} Neither existing \gls{pox} approaches nor \gls{cvm} compartmentalization applies directly. 
Existing \gls{pox} designs prove the execution of individual operations, but depend on features that \glspl{cvm} lack or do not uniformly provide: dedicated hardware that monitors \app and maintains \exec~\cite{apex,asap}, or hardware access control that trusted software configures before \app runs~\cite{slapp,pearts}.
\gls{cvm} compartmentalization runs on server-grade hardware and divides \glspl{cvm} into isolated compartments, but relies on mechanisms specific to one platform (\glspl{vmpl} on \gls{sevsnp}, protection keys on Intel, or Arm \gls{cca} firmware extensions) or on software outside the \gls{tee}   ~\cite{veil,nestedsgx,cabin,sabanic2026wallet,huang2026exia,shelter,caec,tetd,erebor,shi2025serverless}. Moreover, a compartment does not record which operation produced a given output, so code inside the \gls{cvm} can overwrite outputs unobserved.

A \gls{pox} \gls{rot} that loads, measures, and launches \app on \vrf's request, must therefore run inside the \gls{cvm} and rely only on what the \gls{cvm} architecture provides. This is what makes the problem challenging: the measurer shares the \gls{cvm} with \app, the software it measures, while the \gls{vmm} is untrusted and no hardware enforces access control between them. Relying on one vendor's in-\gls{cvm} isolation feature would address this only on that vendor's platform. A design that works across \gls{cvm} platforms therefore requires:
\begin{enumerate*}[leftmargin=*,nosep]
    \item intra-\gls{cvm} isolation between \app and the measurer;
    \item visibility into \app's entries and exits and into every access to its outputs, from which \exec is maintained;
    \item exclusive write access to \exec for the measurer, which updates it throughout \app's run-time;
    \item construction of \gls{pox} evidence that shows \vrf both that \app executed and whether its reported outputs can be trusted.
\end{enumerate*}

\subsection{System and Adversary Model}\label{sec:system}\label{sec:adv}


\textbf{System Model.} We target server-grade processors with \gls{cvm} capabilities. The prover (\prv) is a model provider that serves \gls{ml} inference from a \gls{cvm} hosted by a cloud provider, and the verifier (\vrf) is a client or auditor that sends inference requests and wants assurance that the requested model produced the returned output, e.g., that a request was not routed to a smaller, less capable model. \prv runs an \gls{ml} payload inside a \gls{cvm} managed by a host \gls{vmm}, which creates and configures the \gls{cvm} and controls its \gls{io} access (e.g., network devices).


The \gls{cvm} utilizes virtual firmware that provides minimal functionality needed to launch guest workloads in the \gls{cvm}.
It typically supports \emph{bare-metal containers}: applications that run directly on the firmware or atop a simplified kernel~\cite{TD-Shim-spec,kuvaiskii2024gramine}. We focus on \gls{ml} workloads in the former category, which we call the \emph{payload}, and discuss implications for the latter in Section~\ref{sec:discussion}. We assume the payload consists of multiple \emph{operations} that \vrf may request, each with its own model and code, which may share library code.
We assume a minimal set of baseline protections common across commercial \gls{cvm} architectures: data pages are non-executable, code pages are read-only, return addresses are protected by a shadow stack, and hardware or compiler support provides coarse-grained forward-edge \gls{cfi}. We discuss the specific features of each \gls{cvm} architecture in Section~\ref{sec:formal-versatile}.


\textbf{Adversary Model.} An adversary (\adv) manipulates payload execution to induce unintended behavior in inputs, operations, or outputs, e.g., by modifying data or redirecting control flow to overwrite another operation's outputs. \adv also controls the host environment, including the \gls{vmm} and everything outside the \gls{cvm} (e.g., other \glspl{vm}, storage, and \gls{io} peripherals).
We consider two classes of run-time attacks. \emph{Control-flow hijacking}~\cite{shacham2007geometry,rop,jop} redirects execution to code that would not run in benign execution (e.g., invoking a different operation to write an output). \Gls{ddm}~\cite{non-control-data,data_oriented_exploits} corrupts pointers to write outside the authorized operation (e.g., overwriting inputs, outputs, or execution state).

Side-channel~\cite{wilke2024tdxdown,rauscher2025tdxploit,lee2025t,tdx_sidechannel,mishra2025oops}, denial-of-service, and physical attacks~\cite{tee-fail} are out of scope.
We trust the \gls{cvm} platform's hardware and firmware (e.g., the \gls{tdx} module on Intel \gls{tdx}) to isolate \glspl{cvm} from each other and from the host, and to provide hardware-backed measurement and attestation~\cite{intel_tdx_base_spec,tdx_security_report}. We assume no isolation within a \gls{cvm}, so once \adv executes code in the \gls{cvm} (e.g., by exploiting a payload vulnerability), it can manipulate any unprotected code or data. Since the \adv-controlled \gls{vmm} creates the \gls{cvm}, trust in the \gls{cvm} comes only from remote attestation.


\subsection{Overview of \acron}\label{sec:design:overview}

\acron addresses the challenges discussed earlier and enables \gls{pox} of \gls{ml} workloads deployed in \glspl{cvm}. It has two main modules: a software \emph{instrumentor} and a run-time \emph{monitor}.
We use Intel \gls{tdx} terminology to describe \acron design, as its prototype is implemented on Intel \gls{tdx} (discussed in Section~\ref{sec:evaluation}). As such, the \gls{cvm} is a \gls{td}, its virtual firmware is TD-Shim~\cite{TD-Shim-spec}, and the \gls{tdx} module and \gls{qe} support attestation; Section~\ref{sec:formal-versatile} describes the counterparts on AMD \gls{sevsnp} and Arm \gls{cca}.
Before deployment,
\acron instrumentor processes \app source code (step 0),
which instruments memory writes and indirect calls to produce an instrumented binary $\mathtt{App_{Instr}}$, which is loaded into the \gls{td} (step 1).

\begin{figure}[t]
    \centering
    \includegraphics[width=\columnwidth]{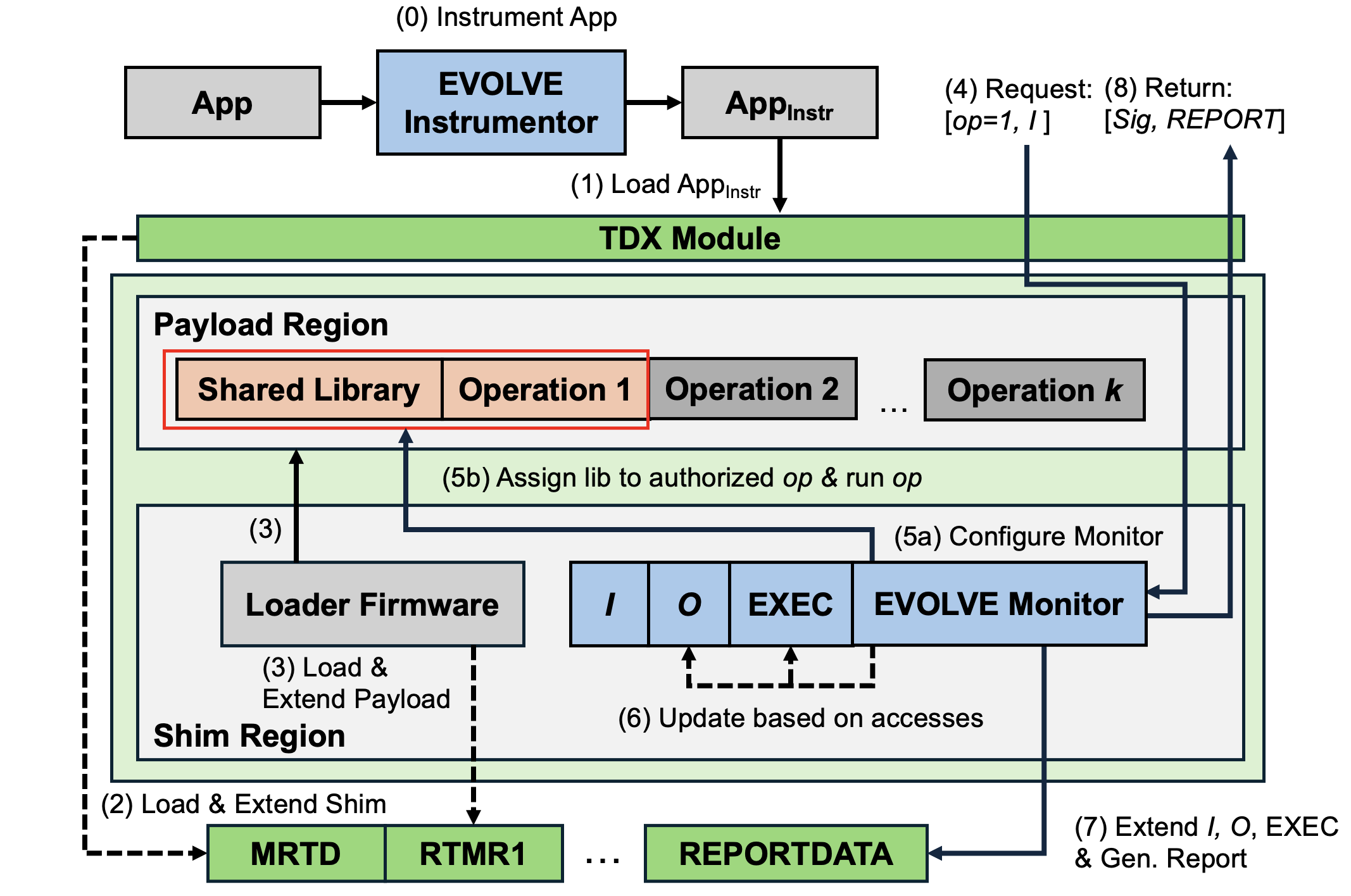}
    \caption{\acron overview. An \gls{ml} \app with multiple operations is instrumented and then loaded and measured into an \acron-enabled \gls{td}. Upon request for operation execution, \acron's monitor assigns the shared library to that operation and tracks access to the output. }
    \label{fig:evolve_overview}
    \vspace{-0.5em}
\end{figure}

\acron splits \gls{td} memory into a \textit{shim region} and a \textit{payload region}, each divided into named \textit{sections} with fixed address ranges set by the linker at build time.

The shim region contains the virtual firmware (e.g., TD-shim), \acron monitor code, and \textit{critical sections} for the input ($I$), the output ($O$), and the \exec.
The \acron-extended shim region requires or adds no trust in the host and is visible to \vrf as it is measured into the \gls{td}'s attestation result, similar to placing a monitor in a privileged guest layer on \gls{sevsnp}~\cite{veil}.
In the payload region, each operation occupies its own \textit{operation section} ($op$). For \gls{ml}, each $op$ maps to one model (e.g., $op_1$ to $\textit{model}_1$).
Code shared across operations, such as an \gls{ml} inference library, is placed once in a \textit{shared section} ($sh$), which is unassigned until it is bound to the authorized operation at run-time.

On startup, the \gls{tdx} module loads and measures the shim region (step 2), and the loader firmware loads and measures the instrumented payload (step 3), binding the payload to the \gls{td}'s initial configuration. The user then sends a request $[op, I]$ with the operation selection and input (step 4). The monitor configures itself and stores the input in $I$ (step 5a), then binds the shared section to the requested $op$ and dispatches it (step 5b). During execution, every instrumented memory write and indirect call passes through the monitor, which updates $O$ and \exec (step 6); accesses from the shared section are attributed to the bound operation. \exec remains 1 if all accesses are authorized and is set to 0 on any violation. At the end, the monitor binds $I$, $O$, and \exec to a hardware-attested report from the \gls{tdx} module (step 7), which is returned to the user as proof that the operation executed without violation (step 8).

\acron addresses the challenges of \gls{pox} beyond embedded systems in two ways. Preprocessing and instrumentation provide intra-\gls{cvm} isolation by defining the shim and payload regions and ensuring exclusive access to the critical sections (discussed further in Section~\ref{sec:design:inst}). The monitor checks memory writes and indirect calls and updates \exec to attest output lineage for the requested operation (discussed further in Section~\ref{sec:design:monitor}).

\section{Instrumentation Module}\label{sec:design:inst}

This section describes the components used in \acron instrumentor. We first discuss the relevant LLVM features, then describe the instrumentation pass, the \gls{ffi} layer, and linking.

\subsection{LLVM Compiler}\label{subsec:llvm}

LLVM~\cite{lattner2004llvm} is a compiler infrastructure built around a typed, low-level \gls{ir} shared across source languages and target architectures. Source programs are translated into this \gls{ir}, which can be transformed or optimized before LLVM's backend lowers it to target-specific machine code. Because all frontends produce this \gls{ir} and all backends consume it, transformations written against it apply regardless of source language or target architecture. This property makes \gls{ir}-level instrumentation a well-established strategy for software security~\cite{ccfi,orenbach2019cosmix,tice2014enforcing,kim2020hardware}, memory safety, and \gls{sfi}~\cite{nagarakatte2009softbound,palit2019mitigating,sehr2010adapting,braden2016leakage}.

LLVM \gls{ir} represents a program as functions of basic blocks in \gls{ssa} form~\cite{llvm_lang_ref}. Three instructions are relevant to our design: \texttt{store}, the main form of memory writes; \texttt{alloca}, which allocates stack space for local variables; and \texttt{call}, which invokes a function directly via a static symbol or indirectly via a function pointer.
LLVM's pass infrastructure allows custom transformations to be inserted into this pipeline. A pass, a self-contained unit that inspects or rewrites \gls{ir}, is implemented as a class with a \texttt{run} method and scheduled by LLVM's pass manager~\cite{llvm_pass_manager}. Passes can be packaged as plugins,
so they do not need to be compiled into LLVM itself. Plugins register at specific extension points in the optimization pipeline, which determine when the custom pass runs relative to other optimizations.

\subsection{Instrumentation Pass}

\begin{figure}[t]
    \centering
    \includegraphics[width=\columnwidth]{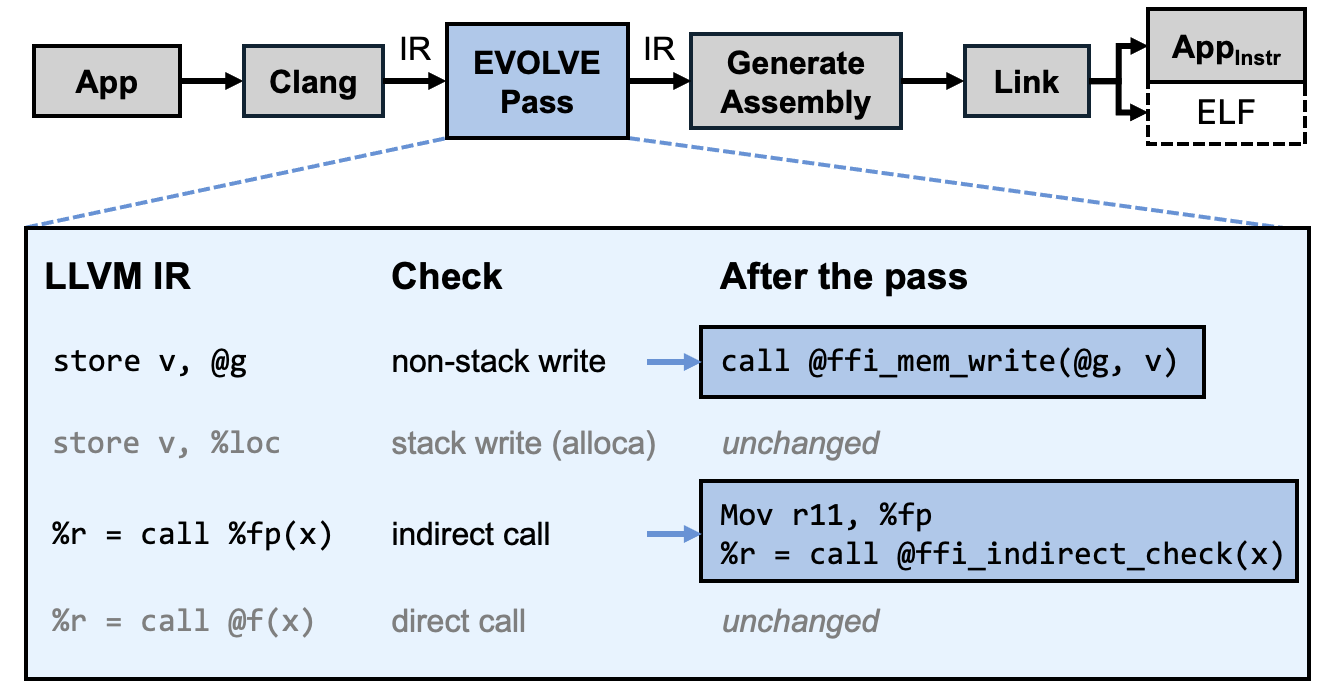}
    \vspace{-1em}
    \caption{\acron instrumentor. An LLVM pass rewrites each non-stack store and indirect call into a call to the monitor's \gls{ffi} layer before the code is compiled to assembly. Stack writes and direct calls are left unchanged.}
    \label{fig:instrumentor}
    \vspace{-0.5em}
\end{figure}

The \acron instrumentor adds a pass to the LLVM pipeline at the \gls{ir} stage, before assembly generation. As shown in Figure~\ref{fig:instrumentor}, the normal pipeline compiles source code to LLVM \gls{ir}, generates assembly, and links the result into the final binary. The pass rewrites the payload's memory writes and indirect calls in between.
The pass handles memory write and indirect call IR.

\textbf{Memory writes.} The pass checks every \texttt{store} in the payload, with two exceptions. First, the monitor's \gls{ffi} functions are excluded via an exact-match allowlist, so the monitor does not intercept its own writes. Second, stores to stack-allocated variables (via \texttt{alloca}) are left unchanged. The stack occupies an address range separate from the critical sections, so an \texttt{alloca}-derived pointer cannot point to $I$, $O$, or any other critical section, and any pointer that can is instrumented.
For each remaining \texttt{store}, the pass replaces it with a call to \texttt{ffi\_mem\_write} with the destination address and value, so the monitor can enforce access control and update the \gls{pox} evidence (discussed in Section~\ref{sec:design:monitor}).


\textbf{Indirect calls.} The pass similarly checks every \texttt{call} instruction. Direct calls to a fixed function are left unchanged. For an indirect call, whose target is a function pointer unknown at compile time, the pass places the original target in a reserved scratch register (e.g., \texttt{r11} in Figure~\ref{fig:instrumentor}) and replaces the callee with \texttt{ffi\_indirect\_check} in the monitor's \gls{ffi} layer, which validates the target before the call proceeds. As a result, no indirect call remains in the instrumented binary. The modified \gls{ir} then continues through assembly generation and linking, leaving the payload source code unchanged. Writes inside indirect-call targets are also instrumented like all other writes.

\textbf{\gls{ffi} layer.} The instrumented payload is linked with the monitor into an \gls{elf} binary. To avoid coupling the payload to the monitor's internal implementation, the payload calls only a set of C-callable \gls{ffi} functions (e.g., \texttt{ffi\_mem\_write}, \texttt{ffi\_indirect\_check}). Each \gls{ffi} function captures the caller's return address as proof of call origin and passes it to the corresponding monitor function for checking (Section~\ref{sec:design:monitor}), so the monitor's internals remain isolated from the payload.

A custom linker script places each component into its designated section. Code shared by multiple operations (e.g., an \gls{ml} inference library) is instrumented and compiled once into a shared section, which every operation calls instead of carrying its own copy. The resulting \gls{elf} binary is packaged with the loader firmware image.

\section{Monitor}\label{sec:design:monitor}


This section describes \acron monitor, the \gls{pox} \gls{rot} inside the \gls{cvm}. We first describe its components, namely the critical data it maintains and its structure, and then its behavior before, during, and after execution. Finally, we discuss the verified primitives it relies on and the security properties they guarantee.

\subsection{Components and Run-time Behavior}\label{sec:design:monitor_comp}

\subsubsection{Components}

\acron monitor, shown in Figure~\ref{fig:monitor}, consists of the critical data it maintains ($I$, $O$, and \exec), an enforcement layer, and a library of verified primitives, which we describe next; the payload reaches it only through the \gls{ffi} layer (recall Section~\ref{sec:design:inst}).

\textbf{Critical Data.}\label{sec:design:monitor_comp:meta} \acron tracks three values to record what was executed and whether it executed correctly. $I$ holds the input sent by the user, such as the input features and model choice, and is written by the monitor before execution. $O$ holds the operation's result (e.g., the model's prediction) and must only be written by the requested operation. \exec records that the requested operation started and completed and whether a policy violation occurred:
\begin{enumerate}[nosep,leftmargin=*]
    \item A payload write targets a critical section ($I$, monitor, or \exec).
    \item An unauthorized caller (e.g., an unrequested operation) writes to $O$.
    \item An indirect call targets an address in the shim region ($I$, $O$, monitor, or \exec) other than an allowed \gls{ffi} entry point.
    \item An indirect call targets a non-authorized operation, or the shared section while it is unbound.
\end{enumerate}

\textbf{Composition of \exec.} \exec comprises four flags  writable only by the monitor, each corresponding to a requirement for \acron's \gls{pox}:
\begin{itemize}[nosep,leftmargin=*]
    \item \execsub{B} (begin): initially 0, set to 1 when the monitor begins the requested operation; it stays 0 if the request is invalid, and the operation does not run.
    \item \execsub{F} (finish): initially 0, set to 1 when the operation returns.
    \item \execsub{M} (memory writes): initially 1, set to 0 on any memory-write violation previously discussed.
    \item \execsub{I} (indirect calls): initially 1, set to 0 on any indirect-call violation previously discussed.
\end{itemize}
At the end of execution, $I$, $O$, and \exec are packed into \texttt{REPORTDATA} of the \gls{tdx} attestation report; the evidence is valid only if all flags in \exec are set to 1.

\textbf{Monitor structure.} The monitor consists of two parts: a \textit{enforcement layer} and a library of \textit{verified primitives}, formally verified functions that perform the monitor's security checks. The enforcement layer integrates with the virtual firmware (e.g., TD-Shim). It exposes the \gls{ffi} entry points, reads section boundaries from the binary, maintains the monitor state (lookup table, authorized operation and bound shared section, and \exec flags), and calls the \gls{tdx} module to generate the report. Instead of deciding whether an access is allowed, it passes the relevant inputs (e.g., destination and caller addresses and section bounds) to the primitives, described next, and acts on the result.


The primitives fall into six categories, each corresponding to one security property (outlined in Section~\ref{sec:sec_properties}): \textit{range checks} map an address to its section~[\ref{prop:range}]; \textit{bounded name} operations compare and copy section names~[\ref{prop:name}]; \textit{flag operations} compose \exec~[\ref{prop:flags}]; \textit{choice operations} validate the requested operation~[\ref{prop:choice}]; \textit{report operations} pack $I$, $O$, and \exec into \texttt{REPORTDATA}~[\ref{prop:report}]; and \textit{share checks} attribute shared-section accesses to the bound operation~[\ref{prop:share}]. 

\subsubsection{Run-time Behavior}
\begin{figure}[t]
    \centering
    \includegraphics[width=\columnwidth]{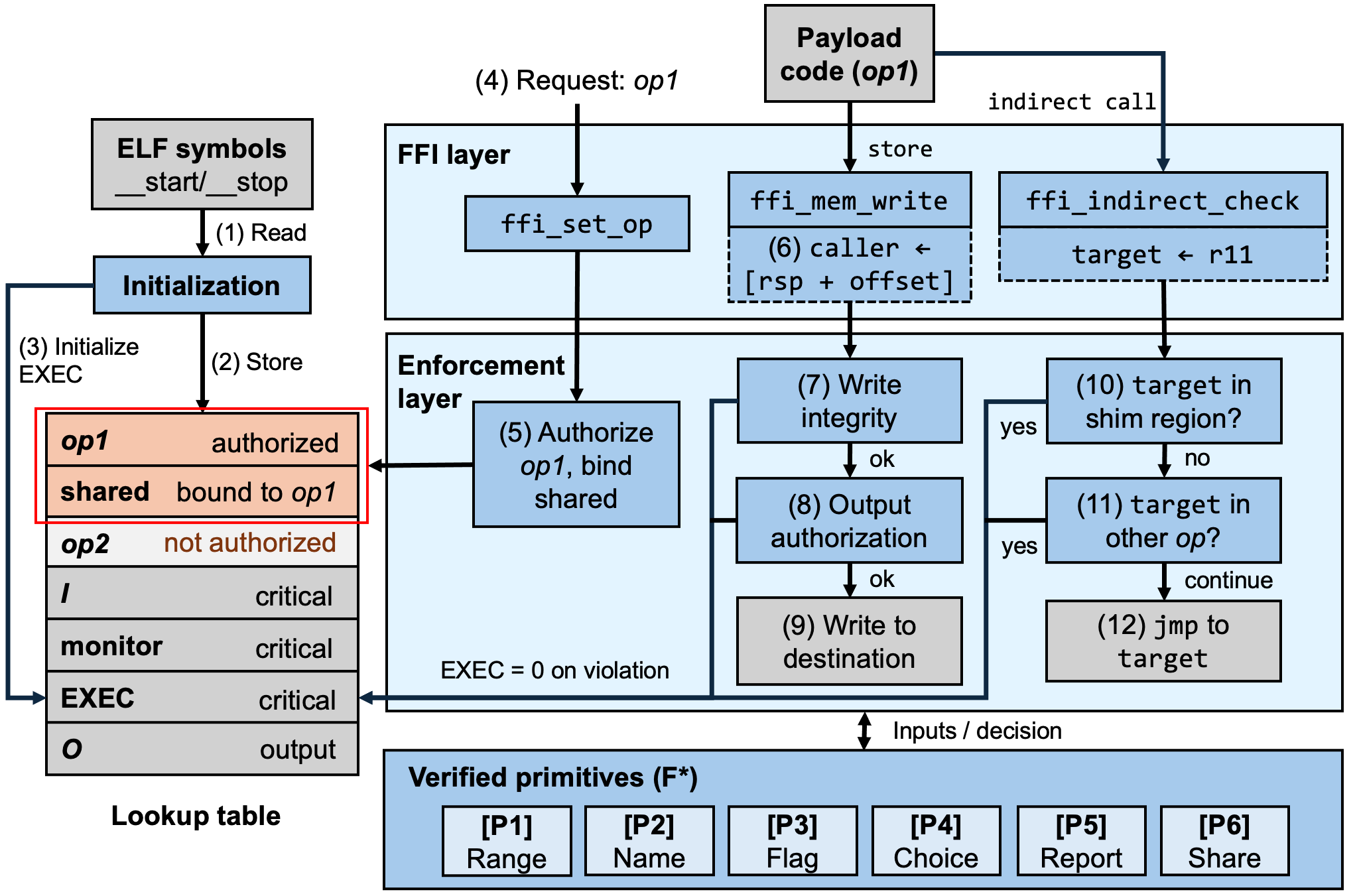}
    \caption{\acron monitor. At startup, the monitor registers each section's bounds in a lookup table (1--3). On a request, it authorizes the chosen operation and binds the shared section to it (4--5). It then checks each memory write (6--9) and indirect call (10--12) from the payload using the verified primitives. The shared section starts unbound and is bound to the authorized operation in step (5), after which it is treated as part of that operation. }
    \label{fig:monitor}
    \vspace{-0.5em}
\end{figure}

The monitor operates in three phases: initialization at startup, enforcement during execution, and attestation after execution. Figure~\ref{fig:monitor} illustrates these phases, and the step numbers below refer to the figure.

\textbf{Initialization.} Before any payload code runs, the monitor establishes the payload memory map and prepares \exec. The monitor is placed in its own section in the shim region, isolating it from the payload at the binary level. As shown in Figure~\ref{fig:monitor}, the monitor first reads each section's start and end addresses, embedded by the linker at build time, from the binary (step 1). It then registers them in a lookup table (step 2), copying each section name into a fixed-size field and matching names only by bounded comparison [\ref{prop:name}]. The table holds $I$, the monitor, \exec, $O$, and one section per operation (e.g., \texttt{model1} and \texttt{model2} in the model-routing scenario). The shared section is also registered, initially bound to no operation. The monitor then initializes \execsub{M} and \execsub{I} to 1 (step 3). Only code in the monitor section can write to \exec [\ref{prop:range}], so the payload cannot manipulate it.

Next, the monitor authorizes the single operation that the user requests. Then, the user sends a request (step 4) specifying the operation (e.g., $op_1$), which the monitor receives through \texttt{ffi\_set\_op}. The monitor first validates the requested operation [\ref{prop:choice}]; an invalid choice sets \execsub{B} to 0, and the operation does not run. Otherwise, the monitor marks $op_1$ as authorized and all other operations as not authorized (step 5), so that only $op_1$ may write to $O$ during this execution. It also binds the shared section to $op_1$ [\ref{prop:share}], so shared section code is treated as part of $op_1$ for the rest of the execution, and sets \execsub{B} to 1.


\textbf{During execution.} During execution, the monitor checks each write and indirect call of the payload against the memory map and the authorized operation, and records any violation in \exec. The payload runs $op$ using $I$. The instrumentor routes every non-stack memory write and every indirect call through the monitor via the \gls{ffi} layer, such as the store and indirect call made by $op_1$'s code in Figure~\ref{fig:monitor}.

The monitor enforces two properties on each memory write. First, \textit{write integrity}, which ensures that the payload never modifies a critical section ($I$, the monitor, or \exec):
when a write arrives via \texttt{ffi\_mem\_write}, the \gls{ffi} layer reads the return address of the call (step 6)
%
from the stack (\texttt{[rsp + offset]} on x86), 
identifying the payload code location that invoked \texttt{ffi\_mem\_write}. The \gls{ffi} layer passes this caller address, along with the destination address, to the monitor.
The monitor then checks whether the destination falls within a critical section ($I$, monitor, or \exec) using the lookup table [\ref{prop:range}] (step 7). If so, the write is blocked and \execsub{M} is set to 0.


Second, \textit{output authorization}, which tracks whether only the requested operation writes $O$. If the destination is $O$, the monitor checks (step 8) whether the caller falls within the authorized operation section set in step (5) [\ref{prop:range}]. A caller in the shared section is also authorized, since the section is bound to $op_1$ [\ref{prop:share}]. If the caller lies anywhere else (e.g., in $op_2$), the write is blocked and \execsub{M} is set to 0. Writes that pass both checks proceed to the destination address (step 9). Any violation is recorded in \exec and is visible to the verifier at attestation time.

For indirect calls, the monitor enforces coarse-grained \gls{cfi}. \texttt{ffi\_indirect\_check} passes the call target to the monitor (step 10), which checks whether it lies within the shim region [\ref{prop:range}]. Such calls are blocked and set \execsub{I} to 0 unless they target an allowed \gls{ffi} entry point, preventing access to internal monitor functions (e.g., one that sets an \exec flag)\footnote{Only entry points need to be checked because Intel \gls{cet}'s \gls{ibt} is assumed to be in place (recall Section~\ref{sec:adv}) and it prevents indirect calls and jumps to arbitrary offsets within functions~\cite{intel-cet}.}. The monitor then checks whether the target lies in an unauthorized operation, or in the shared section while it is not bound to the authorized operation [\ref{prop:share}] (step 11). If so, it then sets \execsub{I} to 0. Calls not blocked in step (10) then proceed to their target (step 12).

\textbf{After execution.} When the operation returns, the monitor sets \execsub{F} to 1 and unbinds the shared section [\ref{prop:share}], so it no longer counts as part of any operation. The monitor then checks that all four flags are exactly 1 [\ref{prop:flags}], confirming that no violation was detected. It packs $I$, $O$, and \exec into \texttt{REPORTDATA} [\ref{prop:report}] and requests a hardware-attested report from the \gls{tdx} module, which \vrf can verify remotely (a \acron-based \gls{pox} protocol is outlined in Appendix~\ref{sec:design:protocol}).


\subsection{Security Properties}\label{sec:sec_properties}


Each category of verified primitives upholds one of the following security properties of \acron monitor; Section~\ref{sec:formal} gives their contracts, and Section~\ref{sec:intuition} explains how the contracts establish each property.
\begin{enumerate}[nosep,label=\textbf{[P\arabic*]},ref=P\arabic*]
    \item\label{prop:range} \textbf{Sound and complete containment}: section membership is an exact interval check $\mathit{start} \leq a < \mathit{end}$ between an address $a$ and a memory range $\mathit{(start, end)}$, and each reported match/mismatch is real.
    \item\label{prop:name} \textbf{Name integrity}: names are compared as bounded C strings that never read past either's bounds. Stored names are truncated to 31 bytes and NUL-padded in a 32-byte field, leaving no stale bytes. Lookup returns the first match, so a later attacker-supplied duplicate cannot shadow an earlier name.
    \item\label{prop:flags} \textbf{Fail-closed \exec flags}: execution succeeds only if all four \exec flags equal exactly $1$. Any other value counts as unset and one unset flag fails the \gls{pox} result.
    \item\label{prop:choice} \textbf{Operation-choice authorization}: a single routine converts a choice integer to an operation index, only for the legal range $1 \ldots \mathit{max\_choice}$, for a upper bound $\mathit{max\_choice}$ determined at compile time. Out-of-range choices are denied regardless of the registration table, and in-range choices are accepted exactly when their operation section is registered.
    \item\label{prop:report} \textbf{Faithful report packing}: $I$, $O$, and the four \exec flags are written to fixed, non-overlapping fields of \texttt{REPORTDATA} with no other byte modified, and a failed flag never appears as passing.
    \item\label{prop:share} \textbf{Shared-section confinement}: accesses from the shared section are authorized only when it is bound to the authorized operation during its execution, and never when it is unbound, bound to another operation, or used outside an execution.
\end{enumerate}

\subsection{Formally Verified Primitives}\label{sec:formal}

This section presents each monitor primitive as a \emph{contract}, which consists of a precondition that must hold at the call and a postcondition that the primitive guarantees on return. Each primitive satisfies its postcondition for every input that meets its precondition. Thus, the correctness of the monitor depends on the correctness of these specifications. Section~\ref{sec:intuition} describes how we verify the contracts with F*~\cite{fstar}, extract them to C, and how they establish properties \ref{prop:range}--\ref{prop:share}. Throughout this section, $\mathit{live}(b)$ means that buffer $b$ is valid and accessible. The condition $h_0 = h_1$ means that the call leaves the heap unchanged, where $h_0$ and $h_1$ denote the heaps before and after the call, respectively.

\subsubsection{Membership and Lookup checks}

\begin{figure}[t]
\centering
\fbox{
    \specsize
    \parbox{0.95\columnwidth}{

        \contract{con:index-of-range}{$r \leftarrow$\code{index_of_range}$(\mathit{starts}, \mathit{ends}, n, a)$}:
        \begin{equation*}\label{eq:contract-range-lookup}
        \begin{aligned}
            \text{requires: }\ &\mathit{live}(\mathit{starts}) \land \mathit{live}(\mathit{ends})
              \land (|\mathit{starts}| = |\mathit{ends}| = n) \\
            \text{ensures: }\ &(h_0 = h_1) \land
              \bigl[(r < n)
            \land (a \ge \mathit{starts}[r] \land a < \mathit{ends}[r]) \\
            \land
            (\forall &k.\ 0 \le k < r
               \Rightarrow \neg(a \ge \mathit{starts}[k] \land a < \mathit{ends}[k])\bigr]
            \\
            \lor \bigl[(&r \ge n)  \land \forall k.\ 0 \le k < n
               \Rightarrow \neg(a \ge \mathit{starts}[k] \land a < \mathit{ends}[k])\bigr]
        \end{aligned}
        \end{equation*}
        
    \contract{con:in-any-range}{$\mathit{res}\leftarrow$\code{in_any_range}$(\mathit{starts}, \mathit{ends}, n, a)$}:
    \begin{equation*}\label{eq:contract-any-range}
    \begin{aligned}
        &\text{requires: }\ \mathit{live}(\mathit{starts}) \land \mathit{live}(\mathit{ends})
          \land (|\mathit{starts}| = |\mathit{ends}| = n)\\
        &\text{ensures: }\ (h_0 = h_1) \land
          \bigl(\mathit{res} \Rightarrow
            (\exists j < n.\ (a \ge \mathit{starts}[j] \land a < \mathit{ends}[j]))\bigr)
    \end{aligned}
    \end{equation*}
    }
}
\caption{Contracts for range membership lookup ([\ref{prop:range}]).}
\label{fig:range-contracts}
\vspace{-0.5em}
\end{figure}

Figure~\ref{fig:range-contracts} gives Contracts~\ref{con:index-of-range} and~\ref{con:in-any-range} for range check membership primitive functions. Both functions take the following inputs:
\begin{itemize}[nosep,leftmargin=*]
    \item \textit{starts:} a buffer containing a list of memory range start addresses;
    \item \textit{ends:} a buffer containing a list of memory range end addresses;
    \item \textit{n}: the number of addresses in \textit{starts} and \textit{ends};
    \item \textit{a}: the address that is undergoing the check.
\end{itemize}
Both contracts require the input buffers to have size \textit{n}.

Contract~\ref{con:index-of-range} specifies the function \textcode{index_of_range}, which returns $r$ as the index of the first range containing $a$, or a value $r \geq n$ if no region contains it. 
The postcondition guarantees that the heap remains unchanged ($h_0 = h_1$). If $r < n$, then $a$ belongs to $r$-th range (i.e., $starts[r] \leq a < ends[r]$), and no earlier region contains it. Otherwise, no region contains $a$.

Contract~\ref{con:in-any-range} specifies \textcode{in_any_range}, which returns a boolean \textit{res} indicating whether $a$ belongs to any of the $n$ ranges. Its postcondition guarantees the heap remains unchanged ($h_0 = h_1$) and that \textit{res} is true if and only if some region contains $a$ (i.e., there exists $j < n$ such that $starts[j] \leq a < ends[j]$).

\subsubsection{Name comparison and registration}

\begin{figure}[t]
\centering
\specsize
\fbox{
    \parbox{0.95\columnwidth}{
        \contract{con:cstr-len}{$r \leftarrow$\code{cstr_len}$(b,n)$}
        \begin{specblock}
            \text{requires: }\ &\mathit{live}(b) \land n \le |b|\\
            \text{ensures: }\ &h_0 = h_1 \land r \le n\\
            &\land\ \forall i < r.\ b[i] \neq 0 \land (r < n \Rightarrow b[r] = 0)
        \end{specblock}

        \contract{con:cstr-eq}{$res \leftarrow$\code{cstr_eq}$(a,n_a,b,n_b)$}
        \begin{specblock}
            \text{requires: }\ &\mathit{live}(a) \land \mathit{live}(b) \land n_a \le |a| \land n_b \le |b|\\
            \text{ensures: }\ &h_0 = h_1 \land \bigl[res \Leftrightarrow
                \exists k \le \min(n_a,n_b).\\
            &\quad \bigl(\forall i < k.\ a[i] = b[i] \neq 0\bigr)\\
            &\quad \land (k < n_a \Rightarrow a[k] = 0) \land (k < n_b \Rightarrow b[k] = 0)\bigr]
        \end{specblock}

        \contract{con:name-zero-tail}{$dst' \leftarrow$\code{name_zero_tail}$(dst,k)$}
        \begin{specblock}
            \text{requires: }\ &\mathit{live}(dst) \land |dst| = 32 \land k \le 32\\
            \text{ensures: }\ &\forall i < 32.\ (i < k \Rightarrow dst'[i] = dst[i])\\
            &\quad \land (i \ge k \Rightarrow dst'[i] = 0)
        \end{specblock}

        \specdef{def:entry-matches}{}
        \begin{specblock}
            &\mathit{entry\_matches}(tbl,i,q,q_{\mathit{len}})
                \triangleq \exists k \le \min(32,q_{\mathit{len}}).\\
            &\quad \bigl(\forall j < k.\ tbl[32i+j] = q[j] \neq 0\bigr)\\
            &\quad \land (k < 32 \Rightarrow tbl[32i+k] = 0)\\
            &\quad \land (k < q_{\mathit{len}} \Rightarrow q[k] = 0)
        \end{specblock}

        \contract{con:find-name}{$r \leftarrow$\code{find_name}$(tbl,cnt,q,q_{\mathit{len}})$}
        \begin{specblock}
            \text{requires: }\ &\mathit{live}(tbl) \land \mathit{live}(q)
                \land 32\cdot cnt \le |tbl| \land q_{\mathit{len}} \le |q|\\
            \text{ensures: }\ &h_0 = h_1 \land r \le cnt\\
            &\land (r < cnt \Rightarrow \mathit{entry\_matches}(tbl,r,q,q_{\mathit{len}}))\\
            &\land \forall i < r.\ \neg\mathit{entry\_matches}(tbl,i,q,q_{\mathit{len}})
        \end{specblock}
    }
}
\caption{Contracts for bounded C-string comparison and section name lookup ([\ref{prop:name}]).}
\label{fig:name-contract}
\end{figure}

Figure~\ref{fig:name-contract} presents Contracts~\ref{con:cstr-len}-~\ref{con:find-name} for the bounded C-string and section name lookup primitive functions.

Contract~\ref{con:cstr-len} specifies \textcode{cstr_len}, which takes a buffer $b$ and a bound $n$, then returns the string length $r$. The precondition requires $b$ to be live and $n$ to not exceed $b$'s size. The postcondition guarantees that the heap remains unchanged $(h_0 = h_1$), that $r \leq n$, and that every byte before $r$ is nonzero (i.e., for all $i < r$, $b[i] \neq 0$. Further, if $r < n$, $r$ is the length, and thus the byte at index $r$ must be zero (i.e., the NUL terminator).

Contract~\ref{con:cstr-eq} specifies \textcode{str_eq}, which returns a boolean \textit{res} reflecting whether two input strings $(a, b)$ with bounds $(n_a, n_b)$ are equal. The precondition requires both $(a,b)$ are live and each be no greater than its specified bound. The postcondition guarantees that the heap remains unchanged ($h_0 = h_1)$. It also guarantees that \textit{res} is true if and only if ($a,b$) contain the same bounded C-string. In particular, there must be some index $k < min(n_a,n_b)$ such that all bytes before $k$ are equal and nonzero. If $k$ is within either bound, the byte at index $k$ must also be zero.

Contract~\ref{con:name-zero-tail} specifies \textcode{name_zero_tail} which takes a destination buffer \textit{dst} and an index \textit{k} and returns the updated buffer \textit{dst'}. The precondition requries \textit{dst} to be live, to have size 32, and to satisfy $k < 32$. The postcondition guarantees that \textit{dst'} preserves every byte before index \textit{k} and sets every byte from \textit{k} onward to zero. 

Contract~\ref{con:find-name}'s postcondition is simplified by defining a predicate \textit{entry\_matches} in Definition 1. It operates over a table of entries \textit{tbl}, an index \textit{i}, a query \textit{q}, and a query length \textit{$q_{len}$}. 
It states that table entry \textit{i} matches \textit{q} when both contain the same nonzero bytes before some index \textit{k}, with index \textit{k} not exceeding 32 or \textit{$q_{len}$}. If \textit{$k<32$}, the table entry must contain a zero byte at index \textit{k}. If $k<q_{len}$, the query must also contain a zero byte at that index.

Contract~\ref{con:find-name} specifies \textcode{find_name}, takes the same inputs and returns an index \textit{r} at which the query name \textit{q} exists. It requires \textit{tbl} and \textit{q} to be live, \textit{tbl} to contain \textit{cnt} entries of 32 bytes each, and \textit{q} not exceed its bound \textit{$q_{len}$}.
It guarantees that the heap remains unchanged ($h_0 = h_1$) and that $r \leq cnt$. If $r < cnt$, it ensures entry $r$ matches $q$ according to Definition 1. It also guarantees that no entry before \textit{r} matches \textit{q}. Thus, \textit{r} identifies the first matching entry or equals \textit{cnt} when no entry matches.

\subsubsection{\exec flag verdicts}

\begin{figure}[t]
\centering
\specsize
\fbox{
    \parbox{0.95\columnwidth}{

        \contract{con:all_set4}{$res \leftarrow$\code{all_set4}$(x)$}
        \begin{specblock}
            \text{requires: }\ &\mathit{live}(x) \land 7 \le |x|\\
            \text{ensures: }\ &h_0 = h_1 \land
                \bigl(res \Longleftrightarrow \forall i \in [3,7).\ x[i] = 1\bigr)
        \end{specblock}

        \contract{con:flag_set_at}{$res \leftarrow$\code{flag_set_at}$(x,\mathit{off})$}
        \begin{specblock}
            \text{requires: }\ &\mathit{live}(x) \land \mathit{off} < |x|\\
            \text{ensures: }\ &h_0 = h_1 \land
                \bigl(res \Longleftrightarrow x[\mathit{off}] = 1\bigr)
        \end{specblock}
    }
}
\caption{Contracts for setting \exec flags ([\ref{prop:flags}]).}
\label{fig:contract-flags}
\vspace{-0.5em}
\end{figure}

Figure~\ref{fig:contract-flags} shows Contracts~\ref{con:all_set4} and~\ref{con:flag_set_at} of the verified primitive functions for management of \exec flags. 
Contract~\ref{con:all_set4} shows the function \textcode{all_set4}, which returns a boolean \textit{res} determining when all \exec flags are set. It takes as input a buffer \textit{x} holding the flags. Its precondition requires $x$ is holding at least $7$ elements, corresponding to the data extended into the report by \acron (described in Sec.~\ref{sec:packed-data}).
As a postcondition, it ensures no heap state change occurs ($h_0 = h_1)$ and \textit{res} is true if all elements in $x$ that are expected to hold \exec flags (3-6) are set to 1.
Contract~\ref{con:flag_set_at} shows the function \textcode{flag_set_at} which returns boolean \textit{res} staring whether the \exec flag at offset \textit{off} in \textit{x} is set. It requires \textit{x} is live and \textit{$\mathit{off} < x$}. It ensures no heap state change occurs ($h_0 = h_1)$ and \textit{res} is true only when \textit{$x[\mathit{off}]=1$}.

\subsubsection{Model-operation Choice Authorization}

\begin{figure}[t]
\centering
\specsize
\fbox{
    \parbox{0.95\columnwidth}{
        \contract{con:choice_auth}{$res \leftarrow$\code{choice_authorized}$(reg,c)$}
        \begin{specblock}
            \text{requires: }\ &\mathit{live}(reg) \land |reg| = \mathit{max\_choice}\\
            \text{ensures: }\ &h_0 = h_1 \land \bigl[res \Leftrightarrow
                (1 \le c \le \mathit{max\_choice}) \land reg[c-1]\bigr]
        \end{specblock}
    }
}
\caption{Contract for model-choice authorization ([\ref{prop:choice}]).}
\label{fig:choice-contract}
\end{figure}

The monitor determines the valid operation from an input integer and refuses to run when the section-table does not hold registered section. Figure~\ref{fig:choice-contract} presents Contract 10 for the primitive function \textcode{choice_authorized}.
It returns a boolean \textit{res} reflecting whether the input choice \textit{c} is authorized based on the set of registers sections statuses in the buffer \textit{reg}. 
It requires \textit{reg} is a live buffer and that is has a size equal to the maximum number of choices (\textit{$max\_choice$}). 
It ensures no heap state change occurs ($h_0 = h_1$) and that \textit{res} is true only when $c$ is a valid choice (i.e., $1 \leq c \leq max\_choice$) and that the section $c$ is indeed registered ($reg[c-1]$ is true).

\subsubsection{Packed report data}\label{sec:packed-data}

\begin{figure}[t]
\centering
\specsize
\fbox{
    \parbox{0.95\columnwidth}{
        \contract{con:pack_report}{$w' \leftarrow$\code{pack_report}$(w,ir_0,ir_1,or,b,f,m,i)$}
        \begin{specblock}
            \text{requires: }\ &\mathit{live}(w) \land 7 \le |w|\\
            \text{ensures: }\ &w' = \langle ir_0, ir_1, or, b, f, m, i \rangle
                \mathbin{\|} w[7..|w|]
        \end{specblock}

        \contract{con:unpack_flags}{$res \leftarrow$\code{unpack_flags}$(w)$}
        \begin{specblock}
            \text{requires: }\ &\mathit{live}(w) \land 7 \le |w|\\
            \text{ensures: }\ &h_0 = h_1 \land \bigl[res \Leftrightarrow
                (w[3]=1 \land w[4]=1\\
            &\quad \land w[5]=1 \land w[6]=1)\bigr]
        \end{specblock}
    }
}
\caption{Contracts for packed report data ([\ref{prop:report}]).}
\label{fig:report-contract}
\vspace{-0.5em}
\end{figure}

Figure~\ref{fig:report-contract} shows Contracts 11-12 for primitive functions that construct report data used in \gls{pox} evidence.
Contract 11 shows \textcode{pack_report}, the function to update the packed report data. It updates the buffer $w$, which becomes $w'$ in the next state. It takes as input the input regions ($ir_0$, $ir_1$), the output region $or$, and the four exec flags from Section~\ref{sec:design:monitor_comp} ($b,f,m,i$).
It requires $w$ is live and has at least seven slots.
It ensures $w'$ contains all of the inputs in order in the first seven buffers. If $w$ is larger than seven slots, the upper slots of $w'$ are equivalent to $w$.

Contract 12 shows \textcode{unpack_flags} which unpacks all \exec flags into a single boolean result \textit{res} that is returned. It takes report data buffer \textit{w} as input.
It requires the same precondition as Contract 11.
It ensures no heap state change, and returns \textit{res} that is true only when all flags are set (located at indices 3 to 6).

\subsubsection{Authorizing shared code}

\begin{figure}[t]
\centering
\specsize
\fbox{
    \parbox{0.95\columnwidth}{
        \contract{con:in_bound_share}{$res \leftarrow$\code{in_bound_share}$(\mathit{in_{ex}}, \mathit{own}, c, a, \mathit{ss}, \mathit{se})$}
        \begin{specblock}
            &\mathit{res} \Longleftrightarrow \mathit{in_{ex}} \land \mathit{own} \neq 0
                \land \mathit{own} = c \land a \ge \mathit{ss} \land a < \mathit{se}
        \end{specblock}

        \contract{con:write_auth}{$res \leftarrow$\code{write_auth}$(\mathit{in_{ex}}, \mathit{own}, c, a, \mathit{st}, \mathit{en}, n, \mathit{ss}, \mathit{se})$}
        \begin{specblock}
            \text{requires: }\ &\mathit{live}(\mathit{st}) \land \mathit{live}(\mathit{en})
                \land |\mathit{st}| = |\mathit{en}| = n\\
            \text{ensures: }\ &h_0 = h_1 \land \bigl[\mathit{res} \Leftrightarrow \mathit{in_{ex}} \land \bigl(\\
            &\quad \exists j < n.\ c = j+1 \land a \ge \mathit{st}[j] \land a < \mathit{en}[j]\\
            &\quad \lor \bigl(\mathit{own} \neq 0 \land \mathit{own} = c
                \land a \ge \mathit{ss} \land a < \mathit{se}\bigr)\bigr)\bigr]
        \end{specblock}

        \contract{con:call_unauth_at}{$res \leftarrow$\code{call_unauth_at}$(\mathit{in_{ex}}, \mathit{own}, c, \mathit{tgt}, n, cs, \mathit{ss}, \mathit{se})$}
        \begin{specblock}
            &\mathit{res} \Longleftrightarrow
                \bigl(cs < n \Rightarrow (\neg\mathit{in_{ex}} \lor c \neq cs+1)\bigr)\\
            &\quad \land \bigl(cs \ge n \Rightarrow
                (\mathit{tgt} \ge \mathit{ss} \land \mathit{tgt} < \mathit{se}\\
            &\qquad \land \neg(\mathit{in_{ex}} \land \mathit{own} \neq 0 \land \mathit{own} = c))\bigr)
        \end{specblock}
    }
}
\caption{Contracts for shared-section authorization ([\ref{prop:share}]).}
\label{fig:share-spec}
\end{figure}

Figure~\ref{fig:share-spec} presents Contracts 12-14 for the primitive functions of shared-section authorization.
Contract 12 shows \textcode{in_bound_share}, that takes the following inputs:
\begin{enumerate*}
    \item \textit{$in_{ex}$}: boolean stating whether execution is occurring;
    \item \textit{own}: the operation that currently owns the shared section;
    \item \textit{c}: the model operation choice;
    \item \textit{a}: the address of interest;
    \item \textit{(ss,se)}: the start/end address of the shared region.
\end{enumerate*}
It is a pure predicate, returning boolean result \textit{res} as true when \textit{$in_{ex}$} is true, \textit{own} equals $c$ and is nonzero, the address of interest is within the shared region (i.e., $ss \leq a < se$). 

Contract 13 specifies \textcode{write_auth}, taking the same previous inputs along with the buffers (\textit{st,en}) containing the start/end addresses of the operation regions. 
It requires that \textit{st} and \textit{en} are live buffers with size \textit{n}.
It ensures no heap state change occurs ($h_0 = h_1)$, and \textit{res} returns as true only when \textit{$in_{ex}$} is true, that either:
\begin{enumerate}[nosep,leftmargin=*]
    \item there exists a registered region that corresponds to the choice $c$ (i.e., that there exists $j<n$ such that $c=j+1$ and $st[j] \leq a < en[j]$); or
    \item \textit{own} equals \textit{c} and is nonzero, and $a$ is in the shared region (i.e., $ss \leq a < se$).
\end{enumerate}

The final contract is \textcode{call_unauth_at} Contract 14, which returns a boolean result \textit{res} stating whether an indirect call to \textit{tgt} is authorized.
It is a pure predicate, returning \textit{res} as true when either the call site is associated with an operation ($cs$) other than the currently executing operation of choice, or the call target ($tgt$) is in the shared region without a matching ownership binding of the operation of choice.

Specifically, if $cs < n$, the call is flagged when it is not within the operation of choice (i.e,. $in_{ex}$ is false) or the call-site index does not correspond to that operation (i.e., $c \neq cs + 1$). When $cs \geq n$, the call is flagged when $tgt$ is outside of the shared region (i.e., $ss \leq tgt < se$) and the shared region is not bound to a currently-executing operation of choice $c$ (i.e., it is false that both $in_{ex}$ is true and $own$ equals a non-zero $c$).


\section{Implementation \& Evaluation}\label{sec:evaluation}

\subsection{Prototype Configuration}

\textbf{Setup.} \acron prototype runs on an Intel \gls{tdx} host with an Intel Xeon Silver 4514Y \gls{cpu} and 512~GB of \gls{ram}. The \gls{td} has 4 vCPUs and 2~GB of \gls{ram}, and QEMU~\cite{qemu} launches the firmware image. The host \glspl{cpu} running the \gls{td} are fixed at their 2.0~GHz base frequency.
We time payload loading, model execution, and quote generation inside the \gls{td} with the \gls{cpu}'s timestamp counter, and build and launch times on the host.
The user inputs are sent to the \gls{td} via a virtual serial port exposed by QEMU as a \gls{tcp} socket. After \acron loads the payload and configures the environment, it waits for a user request. The user requests the model for inference via another serial console, after which \acron executes it and produces the \gls{pox} evidence (as discussed in Section~\ref{sec:design:overview}).



\textbf{Workloads.} Table~\ref{tab:evolve_model} summarizes the models. We train a small (S) and a large (L) configuration of each model type on five datasets from the UC Irvine Machine Learning Repository~\cite{uci_ml_repo}: Breast Cancer Wisconsin Diagnostic (BCWD)~\cite{bcwd_dataset}, ISOLET~\cite{isolet_dataset}, MuskV2~\cite{muskv2_dataset}, Image Segmentation (Image)~\cite{image_seg_dataset}, and Iris~\cite{iris_dataset}. We refer to each model as \texttt{Dataset\_Model\_Size} (e.g., \texttt{BCWD\_MLP\_S}). For \gls{dt}, the configuration refers to the trained structure (number of trees, nodes, leaves, and depth); for \gls{mlp}, to the network architecture (layer sizes and depth).
For \gls{dt}, the S configuration uses the square root of the features as candidates per split, shallow trees (depth bounded by a small constant or the dataset size), and 10 trees.

The L configuration uses more candidate features per split, deeper trees scaled with sample size, and a tree count proportional to input dimensionality.
For example, \texttt{ISOLET\_DT\_S} has 10 trees with 582 nodes, 296 leaves, and depth 5, whereas \texttt{ISOLET\_DT\_L} has 246 trees with 231K nodes, 116K leaves, and an average depth of 20.7. For \gls{mlp}, the S configuration has one hidden layer with one more neuron than the input features, while the L configuration has four hidden layers whose widths decrease linearly from the input dimension to one or two neurons. For example, \texttt{BCWD\_MLP\_S} (30 features) has one hidden layer of 31 neurons, and \texttt{BCWD\_MLP\_L} has layers of sizes [31, 21, 11, 1].
Compiled models have a memory footprint up to 29.17~MB (\texttt{ISOLET\_DT\_L}).

\begin{table}[t]
    \centering
    \caption{Evaluation set includes small (S) and large (L) configurations of \gls{dt} and \gls{mlp} classifiers on five datasets. For \gls{dt}, the configuration specifies the model structure: number of trees, nodes, leaves, and average depth. For \gls{mlp}, it specifies the hidden layer sizes. The last column reports each model's memory footprint.}
    \label{tab:evolve_model}
    \renewcommand{\arraystretch}{1.4}
    \resizebox{\columnwidth}{!}{
    \LARGE
    \begin{tabular}{l!{\vruletab}l!{\vruletab}l!{\vruletab}l}
    \bottomrule

    \toprule
    \textbf{Dataset} & \textbf{Type} & \textbf{Configuration (S, L)} & \textbf{Mem. Size (S, L)} \\
    \midrule
    \multirow{2}{*}{BCWD~\cite{bcwd_dataset}}
      & \gls{dt}  & (10, 318, 164, 5), (12, 356, 184, 6.6)     & 24.53 KB, 28.23 KB \\
      & \gls{mlp} & [31], [31, 21, 11, 1]                      & 12.68 KB, 23.64 KB \\
    \midrule
    \multirow{2}{*}{ISOLET~\cite{isolet_dataset}}
      & \gls{dt}  & (10, 582, 296, 5), (246, 231K, 116K, 20.7) & 45.22 KB, 29.17 MB \\
      & \gls{mlp} & [618], [618, 413, 207, 2]                  & 4.58 MB, 8.35 MB   \\
    \midrule
    \multirow{2}{*}{MuskV2~\cite{muskv2_dataset}}
      & \gls{dt}  & (10, 512, 261, 5), (66, 16.5K, 8.3K, 19.8) & 39.49 KB, 1.86 MB  \\
      & \gls{mlp} & [167], [167, 112, 57, 1]                   & 323.42 KB, 612.73 KB \\
    \midrule
    \multirow{2}{*}{Img.\ Seg.~\cite{image_seg_dataset}}
      & \gls{dt}  & (10, 196, 103, 5), (7, 217, 112, 9)        & 16.39 KB, 20.17 KB \\
      & \gls{mlp} & [20], [20, 14, 8, 2]                       & 7.56 KB, 11.70 KB  \\
    \midrule
    \multirow{2}{*}{Iris~\cite{iris_dataset}}
      & \gls{dt}  & (10, 168, 89, 4.7), (2, 26, 14, 5)         & 13.73 KB, 3.01 KB  \\
      & \gls{mlp} & [5], [5, 4, 3, 2]                          & 1.73 KB, 2.79 KB   \\
    \bottomrule

    \toprule
\end{tabular}
    }
\end{table}

\textbf{Payload layout.} We compile the models into C header files for TD-Shim bare-metal workloads using \texttt{emlearn}~\cite{emlearn}. The S and L models of each type are linked into one payload, each in its own operation section, and the request selects which one runs.
For \gls{mlp}, the shared inference library (layer forward pass, activation functions, and argmax) is compiled once into the shared section. Because this library is generic and receives layer sizes, weights, and biases as model data at run-time, the shared section is identical for every \gls{mlp}, and each operation section holds only the model's entry point and library call. For \texttt{Iris\_MLP\_S} and \texttt{Iris\_MLP\_L}, each operation section is 0.4~KB with one 3.6~KB library in the shared section.
The \gls{dt} models contain self-contained tree and voting logic and do not use the shared section.

For each model, $I$ and $O$ are initialized to 0. At run-time, $I$ is updated based on the user request, and $O$ is updated with the operation's output. All four \exec flags start at 0, and are set to 1 by \acron (as outlined in Section~\ref{sec:design:monitor_comp}).

\subsection{Cost Analysis}
%

To evaluate \acron across \gls{ml} workloads, we use model routing~\cite{shnitzer2023large,guha2024smoothie} as a representative scenario. The router selects between random-forest \gls{dt} and \gls{mlp} classifiers with different levels of complexity. We analyze \acron's costs in two categories: fixed installation costs and run-time performance costs.

\subsubsection{Fixed Costs}
\acron's LLVM pass adds 245~KB to the build system but does not affect the deployed firmware size or the \gls{td} memory footprint. The \acron monitor increases the TD-Shim binary from 221.5~KB to 232.4~KB, a 10.9~KB (4.92\%) increase. This cost reflects the memory allocated to the monitor, which remains active throughout the payload's run-time. In contrast, standard TD-Shim only launches the payload.

\subsubsection{Performance Costs}
Performance costs capture the impact of \acron on building and running the payload. We measure the following:
\begin{itemize}[leftmargin=*,nosep]
    \item \textbf{Baseline:} build and execution time of the payload without \acron.
    The baseline still produces a \gls{tdx} quote of the \gls{td}'s initial state, but no evidence of which operation ran or what wrote its output. For \gls{tdx} overheads on non-\gls{tdx} baseline, we refer readers to~\cite{confidential_llm_tdx_h100}.
    \item \textbf{Build-time overhead:} the additional compile and link time due to incorporating \acron's modules. 
    \item \textbf{Run-time overhead:} additional time to launch the \gls{td} and execute the payload, capturing the cost of routing memory writes and indirect calls through the monitor.
    \item \textbf{Monitor call time:} time the monitor spends performing its checks as total registered operation sections increase.
\end{itemize}

\begin{figure}[t]
\centering
\begin{tikzpicture}
    \begin{axis}[
        width=\columnwidth, height=3.2cm,
        ymin=0, ymax=1.3, ytick={0,0.5,1.0},
        yticklabel style={/pgf/number format/fixed, /pgf/number format/precision=1, /pgf/number format/fixed zerofill},
        ylabel={Norm. build time}, scaled y ticks=false,
        ylabel style={font=\footnotesize}, y tick label style={font=\footnotesize},
        symbolic x coords={BCWD,ISOLET,MuskV2,Image,Iris},
        xtick=data,
        xticklabels={\bverb{BCWD}, \bverb{ISOLET}, \bverb{MuskV2}, \bverb{Image}, \bverb{Iris}},
        x tick label style={font=\footnotesize},
        ybar=0.1pt, bar width=6pt, enlarge x limits=0.12,
        legend style={at={(0.5,1.02)}, anchor=south, legend columns=4, draw=none, font=\scriptsize, /tikz/every even column/.append style={column sep=4pt}},
        area legend,
        error bars/y dir=both, error bars/y explicit,
        error bars/error bar style={black, line width=0.3pt},
    ]
        \addplot+[ybar, fill=\Baseline, draw=black, line width=0.25pt] coordinates {
        (BCWD,1) +- (0,0.036) (ISOLET,1) +- (0,0.040) (MuskV2,1) +- (0,0.041) (Image,1) +- (0,0.042) (Iris,1) +- (0,0.044)};
        \addlegendentry{DT Baseline}
        \addplot+[ybar, fill=\Total, draw=black, line width=0.25pt] coordinates {
        (BCWD,1.158) +- (0,0.063) (ISOLET,1.154) +- (0,0.085) (MuskV2,1.171) +- (0,0.076) (Image,1.157) +- (0,0.073) (Iris,1.150) +- (0,0.075)};
        \addlegendentry{DT EVOLVE}
        \addplot+[ybar, fill=\BaselineG, draw=black, line width=0.25pt] coordinates {
        (BCWD,1) +- (0,0.047) (ISOLET,1) +- (0,0.061) (MuskV2,1) +- (0,0.039) (Image,1) +- (0,0.045) (Iris,1) +- (0,0.048)};
        \addlegendentry{MLP Baseline}
        \addplot+[ybar, fill=\TotalG, draw=black, line width=0.25pt] coordinates {
        (BCWD,1.185) +- (0,0.069) (ISOLET,1.168) +- (0,0.094) (MuskV2,1.166) +- (0,0.071) (Image,1.175) +- (0,0.076) (Iris,1.169) +- (0,0.086)};
        \addlegendentry{MLP EVOLVE}
    \end{axis}
\end{tikzpicture}
\caption{Build time with \acron for \gls{dt} and \gls{mlp} payloads, each containing both the S and L models, normalized to the baseline (mean $\pm$ stdev of 60 builds).}
\label{fig:total_overhead}
\end{figure}

\textbf{Build-time overhead.}
Figure~\ref{fig:total_overhead} shows the build time for model routing payloads (containing both \texttt{S} and \texttt{L}) normalized to the baseline for each model type and dataset.
Build-time overhead is consistent across all five datasets for each model type. For \gls{dt} models, \acron increases build time to 1.15--1.17$\times$ the baseline (25.7--29.2~s to 29.7--33.7~s).
For \gls{mlp} models, the increase is slightly higher at 1.17--1.18$\times$ (25.6--35.3~s to 30.0--41.2~s) 
due to their larger size.
This overhead is purely due to \acron's instrumentation pass on the payload and compilation of virtual firmware containing \acron's monitor.
Build time is a one-time cost per deployment, as the firmware image is built once before it is launched and used for multiple requests.

\begin{figure}[t]
\centering
\begin{tikzpicture}
    \begin{axis}[
        width=0.97\columnwidth, height=4.2cm,
        ymin=0, ymax=8, ytick={0,1,2,3,5,7},
        ylabel={Norm. exec. time}, scaled y ticks=false,
        ylabel style={font=\footnotesize}, y tick label style={font=\footnotesize},
        symbolic x coords={BCWD-S,BCWD-L,ISOLET-S,ISOLET-L,MuskV2-S,MuskV2-L,Image-S,Image-L,Iris-S,Iris-L},
        xtick=data,
        xticklabels={\bverb{BCWD-S}, \bverb{BCWD-L}, \bverb{ISOLET-S}, \bverb{ISOLET-L}, \bverb{MuskV2-S}, \bverb{MuskV2-L}, \bverb{Image-S}, \bverb{Image-L}, \bverb{Iris-S}, \bverb{Iris-L}},
        x tick label style={font=\footnotesize, rotate=30, anchor=north east, inner sep=1pt},
        ybar=0.1pt, bar width=3.5pt, enlarge x limits=0.1,
        legend style={at={(0.5,1.02)}, anchor=south, legend columns=4, draw=none, font=\scriptsize, /tikz/every even column/.append style={column sep=4pt}},
        area legend,
        error bars/y dir=both, error bars/y explicit,
        error bars/error bar style={red!80!black, line width=0.5pt},
        error bars/error mark options={rotate=90, mark size=1.5pt, line width=0.5pt, red!80!black},
    ]
        \addplot+[ybar, fill=\Baseline, draw=black, line width=0.25pt] coordinates {
        (BCWD-S,1) +- (0,0.086) (BCWD-L,1) +- (0,0.078) (ISOLET-S,1) +- (0,0.046) (ISOLET-L,1) +- (0,0.030) (MuskV2-S,1) +- (0,0.081) (MuskV2-L,1) +- (0,0.141) (Image-S,1) +- (0,0.066) (Image-L,1) +- (0,0.078) (Iris-S,1) +- (0,0.079) (Iris-L,1) +- (0,0.072)};
        \addlegendentry{DT Baseline}
        \addplot+[ybar, fill=\Total, draw=black, line width=0.25pt] coordinates {
        (BCWD-S,1.550) +- (0,0.158) (BCWD-L,2.229) +- (0,0.683) (ISOLET-S,1.505) +- (0,0.082) (ISOLET-L,1.002) +- (0,0.040) (MuskV2-S,1.543) +- (0,0.176) (MuskV2-L,1.037) +- (0,0.182) (Image-S,1.873) +- (0,0.193) (Image-L,2.176) +- (0,0.258) (Iris-S,2.218) +- (0,0.301) (Iris-L,2.775) +- (0,0.295)};
        \addlegendentry{DT EVOLVE}
        \addplot+[ybar, fill=\BaselineG, draw=black, line width=0.25pt] coordinates {
        (BCWD-S,1) +- (0,0.077) (BCWD-L,1) +- (0,0.063) (ISOLET-S,1) +- (0,0.007) (ISOLET-L,1) +- (0,0.039) (MuskV2-S,1) +- (0,0.113) (MuskV2-L,1) +- (0,0.079) (Image-S,1) +- (0,0.050) (Image-L,1) +- (0,0.063) (Iris-S,1) +- (0,0.086) (Iris-L,1) +- (0,0.075)};
        \addlegendentry{MLP Baseline}
        \addplot+[ybar, fill=\TotalG, draw=black, line width=0.25pt] coordinates {
        (BCWD-S,3.890) +- (0,0.310) (BCWD-L,6.948) +- (0,0.502) (ISOLET-S,1.194) +- (0,0.009) (ISOLET-L,1.313) +- (0,0.069) (MuskV2-S,1.947) +- (0,0.241) (MuskV2-L,2.716) +- (0,0.224) (Image-S,3.532) +- (0,0.445) (Image-L,6.078) +- (0,0.451) (Iris-S,2.479) +- (0,0.245) (Iris-L,3.916) +- (0,0.321)};
        \addlegendentry{MLP EVOLVE}
    \end{axis}
\end{tikzpicture}
\vspace{-1em}
\caption{Model execution time with \acron, normalized to the baseline (baseline = 1), for \gls{dt} and \gls{mlp} models across five datasets at S and L sizes (mean $\pm$ stdev, $n=30$; one outlier run excluded for the \gls{dt} baseline on Image-S).}
\label{fig:dt_runtime_overhead}
\end{figure}

\textbf{Run-time overhead.} 
Run-time overhead is negligible across all configurations, remaining within $\pm$0.4\% of the baseline (3.04~s on average), except for \texttt{Iris\_MLP\_S}, which incurs 0.89\% overhead due to two slower runs. 
We break down the run-time into three phases: (1) initializing and loading the payload, (2) configuring the monitor and executing the operation, and (3) quote generation. We omit the first phase from the figure because \gls{tdx} initialization dominates it: it takes 2.46~s in every configuration and stays within $\pm$0.4\% of the baseline (1.01\% for \texttt{Iris\_MLP\_S}).


Model execution overhead (shown in Figure~\ref{fig:dt_runtime_overhead}) depends on model type, size, and dataset. For \gls{dt}, \acron increases execution time to 1.50--2.78$\times$ the baseline for all but the two largest trees (only 2--4~$\mu$s per inference) because tree inference requires few memory writes. For the two largest trees (\texttt{ISOLET\_DT\_L} and \texttt{MuskV2\_DT\_L}), the difference stays within one standard deviation.
For \gls{mlp}, \acron increases execution time to 1.19--6.95$\times$ the baseline. The increase is lowest on ISOLET (1.19--1.31$\times$) and highest for the L variants of BCWD and Image (6.08--6.95$\times$). The reason is the balance between computation and monitored writes. Each monitored write adds a fixed cost of about 0.5~$\mu$s (Figure~\ref{fig:call_time_write}), independent of the model.
%
BCWD and Image \glspl{mlp} perform little computation between monitored writes, so the time spent in the monitor exceeds the time spent computing, and their L variants add even more writes. In contrast, ISOLET performs much more computation between writes, so the same monitoring cost is only a small fraction of the total run-time.
We omit quote generation from the figure because it is the same across all configurations, taking 0.020--0.022~ms both with and without \acron. 
This constant cost is due to a single \texttt{TDCALL} after execution that is independent of model size or dataset.

Overall, \acron increases the per-request run-time (i.e., operation execution and quote generation) by an average of 1.07$\times$ $\pm$0.04 for \gls{dt} and 2.13$\times$ $\pm$0.81 for \gls{mlp} relative to the baseline, ranging from no measurable increase for \texttt{ISOLET\_DT\_L} to 3.73$\times$ for \texttt{BCWD\_MLP\_L}; the largest absolute increase is 1.864~ms (1.31$\times$), for \texttt{ISOLET\_MLP\_L}.

\begin{figure}[t]
\centering
\begin{tikzpicture}
\begin{axis}[
    ybar, bar width=0.2cm, width=\linewidth, height=4cm,
    ylabel={Time ($\mu$s)}, ymin=0, ymax=3.5, ytick={0,1,2,3},
    yticklabel style={/pgf/number format/fixed, /pgf/number format/precision=1, /pgf/number format/fixed zerofill},
    xtick=data, symbolic x coords={2,3,4,5,6,7,8,9,10},
    xlabel={Number of operation sections}, enlarge x limits=0.08,
    axis lines*=left, axis line style={black}, tick align=outside,
    tick label style={font=\footnotesize}, label style={font=\footnotesize},
    ymajorgrids, grid style={gray!20},
    legend style={at={(0.02,0.98)}, anchor=north west, font=\footnotesize,
                  draw=none, fill=none},
    legend cell align=left,
]
\addplot[fill=white, draw=black, line width=0.3pt, error bars/.cd, y dir=both, y explicit, error bar style={line width=0.4pt}] coordinates {
    (2, 0.4639) +- (0, 0.0227)
    (3, 0.4721) +- (0, 0.0388)
    (4, 0.5652) +- (0, 0.0299)
    (5, 0.5039) +- (0, 0.0410)
    (6, 0.5324) +- (0, 0.0458)
    (7, 0.6288) +- (0, 0.0418)
    (8, 0.6114) +- (0, 0.0090)
    (9, 0.5486) +- (0, 0.0509)
    (10, 0.5726) +- (0, 0.0272)
};
\addlegendentry{\texttt{ffi\_mem\_write}}
\addplot[fill=\Total, draw=black, line width=0.3pt, error bars/.cd, y dir=both, y explicit, error bar style={line width=0.4pt}] coordinates {
    (2, 0.8166) +- (0, 0.0215)
    (3, 1.0690) +- (0, 0.0609)
    (4, 1.3939) +- (0, 0.0259)
    (5, 1.4799) +- (0, 0.0550)
    (6, 1.8251) +- (0, 0.0530)
    (7, 2.3128) +- (0, 0.0416)
    (8, 2.5865) +- (0, 0.0384)
    (9, 2.7149) +- (0, 0.1939)
    (10, 2.9253) +- (0, 0.0736)
};
\addlegendentry{\texttt{ffi\_indirect\_check}}
\end{axis}
\end{tikzpicture}
\caption{\acron monitor call time for memory writes (\texttt{ffi\_mem\_write}) and indirect call checks (\texttt{ffi\_indirect\_check}) as the number of operation sections increases (mean $\pm$ stdev over 30 runs, run~1 excluded as warm-up). Average of 100 calls per run.}
\label{fig:monitor_call_time}\label{fig:call_time_indirect}\label{fig:call_time_write}
\end{figure}

\textbf{Monitor call time.} 
Figure~\ref{fig:monitor_call_time} further explores the \acron  monitor run-time increase by measuring monitor call time as the number of operation sections $N$ increases from 2 to 10. For memory writes, time per call ranges from 0.46--0.63~$\mu$s without a consistent trend in $N$. This variation is reproducible across repeated runs (within 1\%), suggesting it results from \gls{cpu} cache behavior due to per-run memory layout differences rather than the number of operation sections. The differences are at most 0.17$~\mu$s per call. 
Call time grows linearly for indirect calls from 0.82~$\mu$s at $N$=2 to 2.93~$\mu$s at $N$=10, adding 0.28~$\mu$s per section.

\textbf{Summary.} \acron adds two main costs.
First, building the payload takes 15--18\% longer, but this is a one-time cost per deployment.
Second, \acron increases the per-request run-time by an average of $1.07\times$ $\pm$0.04 for \gls{dt} and $2.13\times$ $\pm$0.81 for \gls{mlp}.
This increase comes from each monitor adding 
0.46--0.64~$\mu$s per write and 0.82--2.93~$\mu$s per indirect call.
In exchange for these costs, \acron provides formal security guarantees (Section~\ref{sec:intuition}) and versatility across \gls{cvm} architectures (Section~\ref{sec:formal-versatile}).

\subsection{Security Analysis}\label{sec:intuition}

\acron extends the TD-Shim~\cite{TD-Shim-spec} virtual firmware with F*-verified policy checks. Each primitive has contracts outlined in Section~\ref{sec:formal}, which \emph{formally guarantee} security properties \ref{prop:range}--\ref{prop:share} from Section~\ref{sec:sec_properties}. We implement the primitives in F*, and extract them to C with KaRaMeL~\cite{karamel}. We link the C code into the Rust firmware through its \gls{ffi}. Our guarantees rely on a correct F* toolchain, C extraction, and C compiler.

[\textbf{\ref{prop:range}]: Sound and complete containment.}
This property requires an exact interval test: $\mathit{start} \leq a < \mathit{end}$. Every reported match must be valid (soundness), and every valid match must be reported (completeness). The check must also use the correct table and strict upper bound. Contract~\ref{con:index-of-range} (\textcode{index_of_range}) and Contract~\ref{con:in-any-range} (\textcode{in_any_range}) establish these conditions. Contract~\ref{con:index-of-range} returns $r < n$ only if range $r$ contains $a$ and no earlier range does. It returns $r \geq n$ only if no range contains $a$. Contract~\ref{con:in-any-range} returns true if and only if some $j < n$ satisfies the same interval test. This equivalence establishes soundness and completeness. The strict upper bound excludes addresses at $\mathit{end}$.

[\textbf{\ref{prop:name}]: Name integrity.}
This property requires bounded C-string comparison, storage of at most 31 name bytes, NUL padding of the remaining bytes in the 32-byte field, and lookup of the first match. Contract~\ref{con:cstr-len} (\textcode{cstr_len}) ensures that the length $r$ does not exceed $n$ and that the routine reads no bytes beyond that bound. Its precondition requires $n$ to fit within the buffer. Contract~\ref{con:cstr-eq} (\textcode{str_eq}) returns true if and only if both inputs hold the same bounded C string. It compares bytes only at indices $k < \min(n_a, n_b)$, so it cannot read beyond either buffer. Contract~\ref{con:name-zero-tail} (\textcode{name_zero_tail}) preserves the first $k$ bytes of the 32-byte field and zeroes all bytes from $k$ onward. Since $k \leq 31$, no stale bytes remain. Contract~\ref{con:find-name} (\textcode{find_name}) returns an index $r$ whose entry matches the query and before which no entry matches. A later duplicate therefore cannot shadow an earlier name.

[\textbf{\ref{prop:flags}]: Fail-closed \exec flags.}
This property requires success only when all four flags equal exactly $1$. Every other value must count as unset, and one unset flag must invalidate the \gls{pox} result. Contract~\ref{con:all_set4} (\textcode{all_set4}) and Contract~\ref{con:flag_set_at} (\textcode{flag_set_at}) establish these conditions. Contract~\ref{con:all_set4} checks exactly four flag slots (indices 3 to 6) and reports success only when each equals $1$. Contract~\ref{con:flag_set_at} applies the same rule to one flag. It returns true only when $x[\mathit{off}] = 1$. Neither routine treats any nonzero value as true. Since the result combines all four flags, one unset flag makes it false. The precondition requires at least seven buffer elements, so all reads remain within bounds.

[\textbf{\ref{prop:choice}]: Operation-choice authorization.}
This property requires one routine to authorize a choice only when $1 \leq c \leq \mathit{max\_choice}$ and the corresponding operation section is registered. Every out-of-range choice must fail, regardless of the registration table. Contract~\ref{con:choice_auth} (\textcode{choice_authorized}) establishes these conditions. The monitor has no other path to an authorization decision. Its postcondition requires two conditions for a true result: $c$ lies in $1 \ldots \mathit{max\_choice}$, and $\mathit{reg}[c-1]$ is true. An out-of-range $c$ therefore always fails. The precondition requires $\mathit{reg}$ to have $\mathit{max\_choice}$ entries, so every valid lookup remains within bounds. For an in-range choice, authorization depends on the registration status of its operation section.

[\textbf{\ref{prop:report}]: Faithful report packing.}
This property requires $I$, $O$, and the four \exec flags to occupy fixed, non-overlapping fields in \texttt{REPORTDATA}. It also requires packing to leave all other bytes unchanged and preserve failed flags. Contract~\ref{con:pack_report} (\textcode{pack_report}) and Contract~\ref{con:unpack_flags} (\textcode{unpack_flags}) establish these conditions. Contract~\ref{con:pack_report} writes $ir_0$, $ir_1$, $or$, and the flags $(b, f, m, i)$ to the first seven slots of $w$ in a fixed order. Each value has its own slot, and every slot beyond the seventh remains unchanged. The routine stores each flag as supplied, so it cannot turn a failed flag into a passing value. Contract~\ref{con:unpack_flags} reads the same flag indices (3 to 6) and returns true only when all four flags equal $1$. Thus, packing and unpacking use the same layout, and unpacking cannot treat a failed flag as passing.

[\textbf{\ref{prop:share}]: Shared-section confinement.}
This property requires access to the shared section to be authorized only during execution and only when the section belongs to the authorized operation. It must deny access when the section is unbound, belongs to another operation, or is accessed outside execution. Contract~\ref{con:in_bound_share} (\textcode{in_bound_share}), Contract~\ref{con:write_auth} (\textcode{write_auth}), and Contract~\ref{con:call_unauth_at} (\textcode{call_unauth_at}) establish these conditions. Contract~\ref{con:in_bound_share} grants access only when $\mathit{in_{ex}}$ holds, $\mathit{own} = c$, $\mathit{own}$ is nonzero, and $ss \leq a < se$. It therefore denies access to an unbound section ($\mathit{own} = 0$), a section bound to another operation ($\mathit{own} \neq c$), or an access outside execution ($\neg\mathit{in_{ex}}$). Contract~\ref{con:write_auth} uses the same binding condition for shared-section writes. It returns true only when $\mathit{in_{ex}}$ holds and either $a$ lies in operation $c$'s registered region or the shared section belongs to $c$. Contract~\ref{con:call_unauth_at} applies the same confinement to indirect calls. It flags calls from sites that belong to another operation and calls into the shared region unless it belongs to the operation currently in execution.

\subsection{Versatility}\label{sec:formal-versatile}

\acron is \emph{versatile} because its modules (an instrumentation pass and monitor extension to guest's virtual firmware) require only features that all major commercial \gls{cvm} platforms provide. Table~\ref{tab:feature-comp} shows that Intel~\gls{tdx}, AMD~\gls{sevsnp}, and Arm~\gls{cca} provide four required features: \gls{dep}, \gls{cfi}, remote attestation, and guest virtual firmware.

\begin{table}[t]
    \centering
    \caption{\acron is versatile across commercial \glspl{cvm} because they all provide four baseline features: \gls{dep}, \gls{cfi}, remote attestation, and guest virtual firmware (GVF).}
    \label{tab:feature-comp}
    \renewcommand{\arraystretch}{1.3}
    \resizebox{\columnwidth}{!}{
    \LARGE
    \begin{tabular}{l!{\vruletab}c!{\vruletab}c!{\vruletab}c!{\vruletab}c}
    \bottomrule

    \toprule
    \textbf{\gls{cvm}} & \textbf{DEP} & \textbf{\gls{cfi}} & \textbf{Remote Attestation} & \textbf{GVF} \\
    \midrule
    \textbf{Intel \gls{tdx}}   & MMU & \gls{cet} (SS + \gls{ibt}) & \gls{tdx} Module + \gls{qe} & TD-Shim          \\
    \textbf{AMD \gls{sevsnp}} & MMU & \gls{cet} SS + compiler     & SP                          & OVMF/Oak stage0  \\
    \textbf{Arm \gls{cca}}     & MMU & PAC + BTI                  & RMM                         & EDK2/direct boot \\
    \bottomrule

    \toprule
    \end{tabular}
    }
    \vspace{-1em}
\end{table}

All three platforms enforce \gls{dep} through the \gls{cpu}'s MMU, which uses execute-permission bits in the guest's page tables. Intel~\gls{tdx} supports \gls{cfi} through Intel \gls{cet}~\cite{intel-cet}, which provides shadow stacks for return addresses and \texttt{ENDBR64} landing pads for indirect calls and jumps. Arm provides \gls{pac} to protect return addresses and \gls{bti} to enforce landing pads for indirect calls and jumps~\cite{arm-pac-bti}. AMD~\gls{sevsnp} uses \gls{cet} shadow stacks~\cite{kvm-svm-shstk} and compiler support for type-based forward-edge \gls{cfi}~\cite{clang-cfi}.

Each platform uses a different root of trust for remote attestation. Intel~\gls{tdx} uses the \gls{tdx} Module and \gls{qe} (discussed in Section~\ref{sec:background}). AMD uses its Secure Processor (SP)~\cite{amd-snp-abi}, while Arm~\gls{cca} uses the \gls{rmm}~\cite{arm-rmm-spec}.
All three platforms also provide virtual firmware that configures and launches the guest payload. Intel~\gls{tdx} uses TD-Shim~\cite{TD-Shim-spec}, which \acron extends. AMD~\gls{sevsnp} uses the Open Virtual Machine Firmware (OVMF)~\cite{ovmf} or minimal firmware such as Oak stage0~\cite{oak-stage0}. The AMD-SP includes either firmware image in the launch digest. Arm~\gls{cca} uses the EFI Development Kit II (EDK2) firmware for Arm virtual machines (ArmVirtQemu)~\cite{edk2-cca}. This firmware contributes to the Realm Initial Measurement (RIM) and loads the payload in the Realm~\cite{arm-rmm-spec}.



\section{Related Work}\label{sec:related_work}

\textbf{\gls{pox} Architectures.} Flicker~\cite{flicker} was an early \gls{pox}-like mechanism. It used late launch and \glspl{tpm} to isolate security-sensitive code from the rest of the system. Recent works extend \gls{pox} to low-end microcontrollers~\cite{asap,apex,pearts,slapp}. APEX~\cite{apex} is a formally verified architecture that provides \gls{pox} for microcontrollers under full software compromise. Its hardware design invalidates \gls{pox} evidence if the executable, output, or monitoring metadata is corrupted or tampered with.
\acron builds on APEX's \gls{pox} definition for controlling \exec value during execution. ASAP~\cite{asap} extends APEX to asynchronous software and supports \gls{pox} for programs that use interrupts. PEARTS~\cite{pearts} introduces \emph{real-time proofs of execution} and preserves execution integrity under real-time scheduling. It also brings APEX's concepts to devices using FreeRTOS. SLAPP~\cite{slapp} extends \gls{pox} to stateful execution and prevents poisoning in federated learning and differential privacy on embedded devices.
Unlike these designs, \acron provides \gls{pox} on server-grade \glspl{cvm} through software. \acron's compiler instrumentation and a monitor in the guest's virtual firmware protect \exec and observe outputs.

\textbf{Attestation of \gls{ml}.} Cryptographic approaches verify \gls{ml} properties without trusted hardware extensions. \Gls{smpc} verifies dataset properties but requires verifier interaction for each attestation~\cite{duddu2024attesting}. Arc combines \gls{smpc} with public commitments to link training and inference~\cite{lycklama2024holding}. \Glspl{zkp} verify training~\cite{zkpot1,zkpot2,zklora}, inference~\cite{zkmlaas,zkcnn,zkllm}, differential privacy~\cite{shamsabadi2024confidentialdpproof,cryptoeprintFranzese}, and fairness~\cite{franzese2024oath,shamsabadi2023confidentialprofitt}. However, their high cost limits deployment. Some require minutes to verify a single training iteration~\cite{zkpot2}. Sentry instead authenticates datasets and models against signed digests as they are loaded into \gls{gpu} memory~\cite{gan2025sentry}, but does not attest how they are used at run-time.
\Gls{tee}-based approaches offer a practical alternative~\cite{duddu2024laminator,attestable_audits,exclavefl,slapp,atlas,chantasantitam2026palm}. Laminator attests classifier properties with Intel \gls{sgx}~\cite{duddu2024laminator}. Schnabl et al. attest code, models, and inference outputs~\cite{attestable_audits}. PAL*M extends property attestation to large generative models on \gls{tee}-enabled \glspl{gpu}~\cite{chantasantitam2026palm}. ExclaveFL attests training inputs and outputs in federated learning~\cite{exclavefl}, while Atlas attests artifact transformations across the \gls{ml} lifecycle~\cite{atlas}. These systems attest \gls{ml} operations but focus on static measurements or specific environments. None provide operation-level \gls{pox} for \gls{ml} workloads in \glspl{cvm}.




\textbf{Intra-\gls{cvm} Isolation.}
Several works build isolated compartments inside \glspl{cvm}. Veil~\cite{veil}, NestedSGX~\cite{nestedsgx}, Cabin~\cite{cabin}, and EXIA~\cite{huang2026exia} use AMD \gls{sevsnp}'s \glspl{vmpl} to host enclaves, confine untrusted programs, or attest external inputs.
Wallet~\cite{sabanic2026wallet} uses them to isolate serverless functions within one \gls{cvm} and attest each invocation's inputs and outputs. SHELTER~\cite{shelter} uses Arm \gls{cca}'s granule protection tables to isolate applications, and CAEC~\cite{caec} extends \gls{cca}'s \gls{rmm} to share memory between realms.
On \gls{tdx}, which has no \glspl{vmpl}, TETD~\cite{tetd} trusts the host \gls{vmm} to hide an agent from the \gls{td} kernel.
Erebor~\cite{erebor} uses \gls{pks} and \gls{cet} to confine a sandbox with an intra-kernel monitor.
COFUNC~\cite{shi2025serverless} uses a microkernel to host isolated serverless functions.
Outside \glspl{cvm}, BlackBox~\cite{van2022blackbox} uses a hypervisor-level monitor to protect containers from an untrusted host.
These systems rely on platform-specific or on software outside our trust model.
TETD trusts the host \gls{vmm}, which the prover may control, and Cabin trusts the guest \gls{os}.
\gls{td} partitioning, the closest \gls{tdx} feature to \glspl{vmpl}, requires a nested \gls{vmm} inside the \gls{td}, multiplying the cost of each \gls{td} exit~\cite{lim2020optimizing}.
These systems also isolate each compartment rather than individual operations. Operation-level isolation would therefore require one compartment per operation. They attest loaded code and, at most, its inputs and outputs, leaving which code produced an output unauditable.
ExclaveFL~\cite{exclavefl} applies this approach to \gls{ml} property attestation, running each federated learning task in a separate \gls{cvm}, and shares these limitations. 
In contrast, \acron runs in the \gls{td}'s guest firmware without platform changes.
It mediates the payload's memory writes and indirect calls, and binds \exec to $I$ and $O$ per operation. Its policy checks use formally verified primitives. This approach follows work formall verification of trusted monitors, such as \gls{cca}'s \gls{rmm}~\cite{li2022design}. However, \acron verifies only these primitives rather than the entire monitor.

\section{Discussion and Future Work}\label{sec:discussion}

\textbf{Incorporating \acron with \gls{ml} Attestation.}
\acron is complementary to existing systems that attest or authenticate \gls{ml} artifacts~\cite{duddu2024laminator,chantasantitam2026palm,atlas,attestable_audits,gan2025sentry}.
These works establish that an \gls{ml} model or dataset is authentic or satisfies a desired property, but they assume that the execution environment correctly performs the computation that uses it. In contrast, \acron proves that a specific operation executed and that the reported output was produced only by that operation, without unauthorized modification. \acron can therefore serve as an execution-attestation layer beneath these systems and become part of their evidence. For example, Sentry~\cite{gan2025sentry} can authenticate a model before it is loaded, and \acron evidence can then establish the integrity of the inference or training operation that uses it. Laminator~\cite{duddu2024laminator} or PAL*M~\cite{chantasantitam2026palm} can attest higher-level properties of the resulting artifacts, and Atlas~\cite{atlas} can record this evidence in their lifecycle provenance. 

\textbf{Adapting to Kernel Payloads.} 
The current design of \acron assumes a bare-metal container payload that executes directly atop the virtual firmware. Virtual firmware (e.g., TD-shim) also support kernel-based deployments, but extending \acron to them requires a more complex design. Instrumenting the stores and indirect calls of both the application and the kernel would incur high overhead, and because the kernel manages application memory and page tables, fixed section boundaries alone can no longer separate operations. One possibility is to draw on enclave models (e.g., Intel \gls{sgx}~\cite{sgx}) to isolate an application from the kernel at run-time while maintaining intra-\gls{cvm} isolation. Ensuring that such a design upholds the same security guarantees requires careful validation, which we leave to future work.

\textbf{Large Generative Models.} The current version of \acron is compatible with \gls{cpu}-based inference workloads that execute entirely within the \gls{cvm}. Future work can extend \acron to large language model inference frameworks such as llama.cpp~\cite{llamacpp}, enabling \gls{pox} for prompt processing and generation. Scaling \acron to modern generative AI workloads, however, requires addressing \gls{gpu} management challenges similar to those in PAL*M~\cite{chantasantitam2026palm}. Although the \gls{gpu} can be directly assigned to the \gls{cvm} as in PAL*M, the NVIDIA drivers that secure communication between the \gls{gpu} and the payload require a kernel, which ties this direction to kernel payloads. Moreover, memory writes performed on the \gls{gpu} are not visible to \acron's instrumentation, so attesting \gls{gpu}-side execution would require extending the monitor's policy to the \gls{gpu}.


\section{Summary}
This work brings proof of execution beyond embedded systems with \acron, an operation-level proof-of-execution system for \gls{ml} workloads in confidential \glspl{vm}. An LLVM instrumentor routes the payload's writes and indirect calls through a run-time monitor in the \gls{td}'s virtual firmware, and the monitor binds its \exec flags to the inputs and outputs in a hardware-attested report. Its policy checks rely on primitives formally verified in F*. A verifier can thus confirm which payload was loaded and that the requested operation produced the output, at negligible end-to-end overhead. 


\ifpublicversion
\section*{Acknowledgments}
This work is supported in part by the Wallenberg Visiting Professor Program, the Natural Sciences and Engineering Research Council of Canada (grant number RGPIN-2026-04826), and the National Cybersecurity Consortium and the Government of Canada (CSIN) (project id 2024-1345).
\fi


\bibliography{references}
\bibliographystyle{IEEEtran}

\appendices

\section{\acron-based PoX Protocol}\label{sec:design:protocol}

Figure~\ref{fig:evolve_protocol} shows the \acron-based \gls{pox} protocol between \vrf and \prv. \prv consists of the untrusted \gls{vmm} and the \gls{qe} on the host machine, the trusted \gls{tdx} module, and the \acron \gls{td}, which is split into the payload region and shim region housing \acron's monitor.

\begin{figure}[t]
\footnotesize
\begin{protocol}{\acron-enabled \gls{pox}}

\noindent\textit{Notation}
\begin{itemize}[nosep, leftmargin=1.2em, label=--]
    \item $I$, $O$: input and output of the payload operation.
    \item $op$, $\chal$: operation choice and challenge nonce chosen by \vrf.
    \item $h_{TD}$: measurement of the \gls{td} firmware (with the monitor) and the instrumented payload.
    \item \exec: four execution flags (\execsub{B}, \execsub{F}, \execsub{M}, \execsub{I}); 
    \item \texttt{TDREPORT} / \texttt{QUOTE}: MAC-ed report from the \gls{tdx} module / signed quote from the \gls{qe}.
\end{itemize}

\protosep

\protophase{Offline}
\begin{enumerate}[label=\arabic*., leftmargin=*, nosep, series=proto]
    \item \texttt{App} is instrumented by an LLVM pass that replaces every non-stack \texttt{store} with \texttt{ffi\_mem\_write} and every indirect call with \texttt{ffi\_indirect\_check}, producing \texttt{App\textsubscript{Instr}}.
\end{enumerate}

\protophase{Load time}
\begin{enumerate}[resume*=proto]
    \item \gls{vmm} specifies the \gls{td} image and invokes \gls{td} creation.
    \item \gls{tdx} module measures the \gls{td} firmware, including the monitor; the firmware then measures \texttt{App\textsubscript{Instr}}, together forming $h_{TD}$, and launches it.
    \item Monitor registers the sections ($I$, $O$, monitor, \exec, one section per operation, and the unbound shared section) and sets \execsub{M} $=$ \execsub{I} $= 1$.
\end{enumerate}

\protophase{Request time}
\begin{enumerate}[resume*=proto]
    \item \vrf sends $\{op, I, \chal\}$ to the \gls{vmm}, which forwards it to the \gls{td}.
    \item Monitor stores $I$ and validates $op$; an invalid $op$ sets \execsub{B} $= 0$ and nothing runs.
    \item Monitor authorizes the section of $op$, binds the shared section to it, sets \execsub{B} $= 1$, and invokes $op$.
\end{enumerate}

\protophase{Execution}
\begin{enumerate}[resume*=proto]
    \item Payload executes the requested operation; every instrumented memory write and indirect call is routed to the monitor.
    \item Monitor checks each event against the policy:
    \begin{itemize}[nosep, leftmargin=1em, label=--]
        \item writes to $I$, the monitor, or \exec, are blocked and clears \execsub{M};
        \item writes to $O$ from outside $op$ or its bound shared section proceed but clear \execsub{M};
        \item an indirect call into the shim region other than an allowed \gls{ffi} entry point is blocked and clears \execsub{I};
        \item an indirect call into another operation or the unbound shared section proceeds but clears \execsub{I};
        \item all other events proceed.
    \end{itemize}
\end{enumerate}

\protophase{After execution}
\begin{enumerate}[resume*=proto]
    \item Monitor sets \execsub{F} $= 1$ when $op$ returns, unbinds the shared section, and requests a \texttt{TDREPORT} with
        \protoeq{\texttt{REPORTDATA} := \{I,\ O,\ \exec,\ \chal\}}
    \item \gls{tdx} module creates the MAC-ed \texttt{TDREPORT}.
    \item \gls{qe} validates the MAC and signs the \texttt{QUOTE}, which the \gls{vmm} sends to \vrf.
    \item \vrf verifies the \texttt{QUOTE} signature and $h_{TD}$, checks \chal, unpacks \exec, and validates $I$ and $O$.
\end{enumerate}

\end{protocol}
\caption{\acron-enabled \gls{pox} protocol.}
\label{fig:evolve_protocol}
\end{figure}

\textbf{Offline phase.} Before any request is made, the application is instrumented with \acron's LLVM pass (recall Section~\ref{sec:design:inst}), producing $\mathtt{App_{Instr}}$, in which every non-stack memory write and indirect call is routed through the monitor (step 1). This step happens once per deployment.

\textbf{Load time.} \gls{vmm} specifies and creates the \gls{td} (step 2). \gls{tdx} module measures the \gls{td} firmware and \acron monitor. Then, the firmware then measures $\mathtt{App_{Instr}}$, which all together form $h_{TD}$ (step 3). At boot, the monitor records the address boundaries of all sections ($I$, $O$, monitor, \exec, one section per operation, and the shared section, initially unbound) and sets \execsub{M} and \execsub{I} to 1 (step 4).

\textbf{Receiving a request.} \vrf initiates the protocol by sending the operation $op$, the input $I$, and an attestation challenge $\chal$ to \prv. The \gls{vmm} forwards the request into the \gls{td} (step 5). \acron's monitor stores $I$ and validates $op$; an invalid $op$ sets \execsub{B} to 0, and nothing runs (step 6). Otherwise, the monitor authorizes the $op$ section (so that only this section may write $O$ during this execution), binds the shared section to it, sets \execsub{B} to 1, and invokes $op$ (step 7).

\textbf{Execution.} The payload runs $op$, and every instrumented memory write and indirect call is routed to the monitor (step 8). The monitor checks each event against the policy (step 9).
Upon writes to critical data, \acron blocks them and clears \execsub{M}. \acron monitor also clears \execsub{M} upon writes to $O$ from outside $op$ or its shared section.
Indirect calls into the shim region other than an allowed \gls{ffi} entry point are blocked and clear \execsub{I}, while indirect calls into another operation or the unbound shared section proceed but clear \execsub{I}. All other events proceed.

\textbf{Generating~\texttt{QUOTE}.} When $op$ returns, the monitor sets \execsub{E} to 1, unbinds the shared section, and requests a \texttt{TDREPORT} from the \gls{tdx} module with \texttt{REPORTDATA} $= \{I, O, \exec, \chal\}$ (step 10). \gls{tdx} module produces the \gls{mac}-ed \texttt{TDREPORT} (step 11). \gls{qe} verifies the \gls{mac} and signs the \texttt{QUOTE}, which the \gls{vmm} sends to \vrf (step 12).

\textbf{Verification.} \vrf verifies the \texttt{QUOTE} signature to confirm that it originated from the expected \gls{td}, and \vrf checks \chal to confirm freshness (step 13). It then compares $h_{TD}$ against a known reference value to confirm that the expected firmware and instrumented binary were used. The final check is \exec.
\vrf unpacks \exec to obtain the four flags. If all are equal to one, \vrf is assured that the requested operation started and completed, that $O$ was written only by the authorized operation, and that no policy violation occurred during execution. If any equal zero, the result is untrustworthy regardless of the value of $O$. 

\end{document}